\documentclass[11pt,a4paper]{article}
\usepackage{jheppub,amsmath,amssymb,slashed,url,bm,textgreek,upgreek,braket}
\usepackage{graphicx}
\usepackage{epstopdf}
\usepackage{amsthm,mathtools}
\usepackage{booktabs,array}
\usepackage{enumitem}
\usepackage{etoolbox}
\hypersetup{hypertexnames=false,pdftitle={Towards Batyrev--Borisov mirrors from a graded Morrison--Plesser parent},pdfauthor={Wei Gu}}
\patchcmd{\maketitle}{\pagenumbering{arabic}}
  {\clearpage\pagenumbering{arabic}}{}
  {\PackageError{JTdraft}{Could not separate the contents from the body}{}}
\patchcmd{\maketitle}
  {\pagestyle{myplain}\clearpage\pagenumbering{arabic}}
  {\clearpage\pagestyle{myplain}\pagenumbering{arabic}}{}
  {} 
\usepackage[T1]{fontenc}
\usepackage{lmodern,microtype,xcolor,pgfplots}
\pgfplotsset{compat=1.18}
\definecolor{linkblue}{HTML}{24526B}
\newcommand{\C}{\mathbb C}
\newcommand{\Z}{\mathbb Z}
\newcommand{\cH}{\mathcal H}

\newcommand{\Jac}{\operatorname{Jac}}
\newcommand{\paperref}[1]{\href{https://arxiv.org/abs/#1}{arXiv:#1}}
\title{Towards Batyrev--Borisov mirrors from a graded Morrison--Plesser parent}
\author{Wei Gu}
\emailAdd{guwei2875@zju.edu.cn}
\affiliation{
  Zhejiang Institute of Modern Physics, School of Physics,\\
  Zhejiang University,
  Hangzhou 310027, China
}

\abstract{We revisit the parent theory proposed by Morrison and Plesser
to understand the Batyrev--Borisov mirror construction from
duality between abelian gauged linear sigma models (GLSMs).
To resolve the sign issue in their construction, we modify
the matter content by introducing Grassmann-odd chiral
superfields and auxiliary gauge groups.
On one side, the additional matter fields become massive
in a Higgs vacuum, while on the other side, their one-loop
contributions give a rational superpotential whose leading
term is the expected mirror superpotential.
For a smooth complete intersection, we show that the
higher-order terms can be absorbed locally into a
redefinition of the normal coordinates even at finite
couplings. In an LG or hybrid phase with the appropriate
IR grading, these terms are irrelevant deformations.
We also study the correspondence between the FI parameters
and the coefficients of the mirror superpotential at the
level of operators. Further evidence comes from the vortex
calculation in the Higgs vacuum, which, in a certain region
of parameters and at fixed ordinary gauge backgrounds,
selects the Coulomb solution used in the mirror calculation.
From these observations, we conjecture that the two GLSMs
associated with a Batyrev--Borisov mirror pair can be obtained
from our extended parent theory and that the corresponding
compact sectors, with the expected discrete gauge projections,
flow to the same IR SCFT under mirror symmetry without an
additional massless sector.
The construction is illustrated by the quintic and the
complete intersection of two cubics in $\mathbb P^5$.}

\begin{document}
\maketitle

\section{Introduction}
\label{sec:introduction}

It has long been expected that the Batyrev--Borisov mirror construction
for toric Calabi--Yau complete intersections~\cite{Batyrev,Borisov,BB,BBcomplete}
can be understood as an infrared duality between abelian gauged linear
sigma models (GLSMs)~\cite{Witten,MP}. In 1995, Morrison and Plesser
(MP) proposed a parent theory for this purpose~\cite{MP}. Their proposal
was motivated by dualities in four-dimensional supersymmetric gauge
theories, including Seiberg--Witten theory~\cite{SW}, where different
gauge descriptions give the same low-energy physics. In the MP proposal,
one uses field rescalings to reach different regimes of the same parent
theory while keeping the physical parameters fixed. Integrating out the
massive fields in these regimes is then expected to give the two mirror
GLSMs. However, a relative sign in the parent action leads to inverse powers of the proposed mirror fields in their reverse calculation, whereas
the expected interaction is polynomial. In this paper, we modify the matter content of the parent theory so that the reverse one-loop calculation gives the expected monomials at leading order.

Let $V$ be the $D$-dimensional toric ambient space. A nef partition divides its toric boundary divisors into $r$ sets. The sum $D_\alpha$ of the divisors in the $\alpha$-th set defines a basepoint-free line
bundle, meaning that its global sections have no common zero. These divisors satisfy $\sum_{\alpha=1}^r D_\alpha=-K_V$. We choose a section $G_\alpha$ of each line bundle, and the common zero locus
of these sections is the complete intersection. In the GLSM, we introduce one field $\mathcal P_\alpha$ for each defining equation. The gauge-invariant superpotential is
$W_A=\sum_\alpha\mathcal P_\alpha G_\alpha$.
For example, the six coordinate hyperplanes of $\mathbb P^5$ can
be divided into two sets of three, which gives $D_1\sim D_2\sim 3H$,
where $H$ is the hyperplane class. $G_1$ and $G_2$ can then be chosen as general cubics in all six homogeneous coordinates.

We use the nef partition and its dual to construct the two GLSMs.
The monomials in their superpotentials are encoded by an integer matrix $P$. Each row of $P$ gives a monomial in the original theory and corresponds to a superfield in the mirror. Each column gives a monomial in the mirror theory and corresponds to a superfield in the original theory. The integral factorization $P=\mathsf S\mathsf T$ also specifies the two gauge groups, including their finite components. We choose the phase of each GLSM by placing its Fayet--Iliopoulos (FI) parameters in a suitable chamber. This construction applies to complete intersections
of arbitrary codimension. The matrix $P$ need not be square, and we do not require that the GLSMs admit an isolated Landau--Ginzburg (LG) phase.

We first ask whether a different choice of mirror variables in the
IR could resolve the sign issue in the MP parent theory. Recall
that Hori and Vafa derived mirror superpotentials by applying
abelian duality to GLSMs and taking vortex contributions into
account~\cite{HV}. For a compact complete intersection and its
associated noncompact Calabi--Yau bundle model, the Hori--Vafa
mirror superpotentials have the same form when written in the
same dual variables. The two models differ, however, in their
choice of fundamental mirror variables. This affects the integration
measure, as one can see by comparing their mirror period integrals.
To obtain the compact unit-state period from the bundle model, one
inserts $\prod_{\beta=1}^{r}\delta_\beta$ into the bundle integral,
where $\delta_\beta$ represents the divisor class of the $\beta$th
defining equation~\cite[eqs.~(7.75)--(7.78)]{HV}.
In~\cite[Section 2 and Appendix B]{GS}, the author and Sharpe proposed a choice of fundamental mirror variable based on the vector R-charge.
For a field of R-charge zero, this variable is the dual field $Y$ itself. In the geometric phase of the compact model, the
$\mathcal P$-field has R-charge two, so its fundamental mirror
variable should be $e^{-Y_p}$.

Returning to the MP parent theory, let $B$ be the monomial in the hatted fields associated with one column of $P$.
The calculation on the proposed mirror side gives a term
proportional to $1/B$, which becomes polynomial if we invert the hatted fields on the dense torus. Whether we should use these variables in the IR also depends on the measure we obtain after integrating out the massive fields in the parent theory. For a complete intersection, we can attack this question via the period integral, since we know which insertion is needed for the period in the compact theory.
In the quintic example, we take a flat holomorphic reference measure
for the hatted fields. On the regular locus, the R-charge factor,
the one-loop determinant and the Jacobian of the inversion then leave
a field-independent multiple of the logarithmic volume form.
Whether the remaining integral over the gauge and matter fields, taken over the appropriate cycle, supplies the insertion in the measure is still unknown.

We also consider a different route by asking what change in the parent theory would give a positive power of $B$ directly in the original hatted fields.
Let us examine the two terms responsible for the inverse power.
Consider an ordinary chiral multiplet of charge one under $U(1)_H$.
In our conventions, its one-loop contribution and the tree-level coupling give the following terms in the effective twisted superpotential,
\[
 -\Sigma_H[\log(\Sigma_H/\mu)-1]+\Sigma_H\log[a/(\mu B)],
\]
where $\Sigma_H$ is the field strength of the $U(1)_H$ vector multiplet and $\mu$ is a reference scale.
The vacuum equation for $\Sigma_H$ gives $\Sigma_{H,*}=a/B$.
At this value, the two logarithmic terms cancel, leaving the inverse monomial $a/B$. Now consider a Grassmann-odd chiral multiplet of the same charge.
Reversing the statistics changes the sign of its one-loop contribution~\cite[eq.~(12)]{GuZou}.
In our graded parent theory, we add two such ghost multiplets to the original parent theory, so the coefficient of $\Sigma_H[\log(\Sigma_H/\mu)-1]$ becomes $-1+2=+1$.

To recover the original GLSM in the other regime of the parent theory, we need these two ghost multiplets to become massive so that we can integrate them out. For each column of $P$, we therefore introduce partners of opposite charge and an auxiliary $U(1)_K$ gauge group. In the vacuum where the charge-one field $S$ has a
nonzero expectation value, the auxiliary gauge group is completely Higgsed and the gauge-invariant superpotential couplings give
masses to the ghost multiplets and their partners. At energies
below these masses, we can integrate out the added matter and apply the MP derivation of the original GLSM in a neighborhood of this vacuum.

In the mirror regime, the partners contribute to the Coulomb branch equations for both $U(1)_H$ and $U(1)_K$. Let $a$ and $\kappa$
denote the holomorphic FI scales of these two gauge factors,
respectively. Solving the two equations on the regular massive
branch gives the holomorphic threshold
\begin{equation}
 F(B)=\frac{\kappa a}{a+\kappa B},\qquad
 F(B)-\kappa=-\frac{\kappa^2}{a}\,B+O(B^2).
 \label{eq:preview}
\end{equation}
The expansion uses the regular continuation of the threshold to
$B=0$. For the $i$-th column of $P$, recall that the coefficient
of the linear term in $B_i$ is $d_i=-\kappa_i^2/a_i$.
Let us choose $\kappa_i=Ld_i$ and $a_i=-L^2d_i$, where
$L$ is finite and nonzero. After subtracting the
field-independent constants, we obtain
\begin{equation}
    W_L=\sum_i\frac{d_iB_i}{1-B_i/L}.
\end{equation}
Each monomial $B_i$ in the dual nef partition contains the
field $\widehat{\mathcal P}_{\beta(i)}$ to the first power and
is independent of the other $\widehat{\mathcal P}$ fields.
Thus, the terms linear in the $\widehat{\mathcal P}$ fields
give the expected superpotential
\begin{equation}
    W_{\rm BB}=\sum_i d_iB_i
    =\sum_{\beta=1}^{r}\widehat{\mathcal P}_\beta
      \widehat G_\beta(\widehat{\mathcal U}),
\end{equation}
where $\widehat{\mathcal U}$ denotes the other matter superfields
on the hatted side.

One may wonder whether the higher-order terms in $W_L$
affect the IR physics. Let us first consider the geometric
phase and assume that the complete intersection
\[
    \widehat Y=\{\widehat G_1=\cdots=\widehat G_r=0\}
\]
is smooth and compact. In this case, we can locally regard
the $\widehat G_\beta$ as independent coordinates in the
normal directions to $\widehat Y$. Since all the additional
terms in $W_L$ are at least quadratic in the
$\widehat{\mathcal P}$ fields, we can absorb them into
these normal coordinates by a holomorphic field
redefinition. In the new coordinates, the superpotential
takes the same form as $W_{\rm BB}$ near $\widehat Y$. This is a local result for the full rational superpotential, without truncating its expansion, and it does not determine the other critical sectors away from $\widehat Y$. Thus, if the kinetic metric is regular, the
$\widehat{\mathcal P}$ fields together with the fields
in the normal directions are massive and can be
integrated out. We are left with the same compact
critical manifold $\widehat Y$. Note that this argument
does not depend on the codimension of the Calabi--Yau
manifold. One can also study these higher-order terms from the
LG or hybrid description whenever such a phase exists.
If $B_i$ and $B_i^n$ have scaling dimensions one and $n$,
respectively, at the IR fixed point, then the terms with
$n>1$ are irrelevant F-term deformations.

Let us now compare the parameters of the two GLSMs.
Recall that $a_i=-L^2d_i$, with the same $L$ for all $i$.
For any $\ell\in\ker_{\mathbb Z}P$, we have
$\sum_i\ell_i=0$ due to the Calabi--Yau condition. Thus,
\begin{equation}
    \prod_i a_i^{\ell_i}=\prod_i d_i^{\ell_i}.
\end{equation}
This is precisely the relation we expect between the
FI parameters of the original GLSM and the coefficients
of the mirror superpotential. On the other side, we can
choose the hatted FI parameters in a similar way.
Note that some of these parameters are redundant due to
field redefinitions. Also, the GLSM description may not
account for all the deformations of the IR SCFT.

To further understand this parameter correspondence, let us
consider the operators coupled to these parameters.
Since $a_i=-L^2d_i$ and $\kappa_i=Ld_i$, varying $\log d_i$ while keeping $L$ fixed amounts to varying both $\log a_i$ and $\log\kappa_i$ by the same amount. Thus, the operator associated with this variation in the parent theory is $\Sigma_{H_i}+\Sigma_{K_i}$.
In the Higgs vacuum where we recover the original GLSM,
this operator locally gives the corresponding FI deformation.
On the mirror side, we can evaluate the same operator by
solving the vacuum equations. The leading field-dependent
term is $d_iB_i$, which is precisely what we obtain by
varying $W_{\rm BB}$ with respect to $\log d_i$.

One may ask whether we have chosen the right solution
of the auxiliary Coulomb equations in the mirror calculation.
To answer this question, we study the vortex contributions
in the $S$-Higgs vacuum while treating the ordinary gauge
multiplets as background fields. For an open region of real
parameters and a fixed ordinary Coulomb scalar, the auxiliary
$U(1)_K$ equation has two real roots. We find that the semiclassical limit of the
vortex sum selects the larger one, which is the solution
we used to obtain the mirror superpotential.

From the above discussion, we expect that our extended
parent theory can be used to understand the general
Batyrev--Borisov mirror construction. More precisely,
for a Batyrev--Borisov mirror pair for which both GLSMs
flow to nonsingular compact SCFTs, we conjecture that
the source-free $S$-Higgs sector and the compact sector
obtained from the reverse calculation describe the same
IR SCFT under mirror symmetry. Here we take the IR limit
while keeping the microscopic couplings fixed. As usual in mirror symmetry, the ordinary chiral sector of
one theory is exchanged with the twisted chiral sector of
the other. We also expect that the additional matter fields
introduced in our parent theory do not give rise to an
extra massless sector in the IR.

In the graded theory, the local calculation also gives a
curvature coupling on the massive branch. We expect that the
measure on the selected compact IR sector agrees with that of
the mirror GLSM, up to a field-independent normalization.
Testing this expectation requires the remaining gauge
integrations, including the finite gauge projection determined
by the integral toric data.

Of course, the mirror pairs we study in this paper are
already known from the Batyrev--Borisov
construction~\cite{Batyrev,Borisov,BB,BBcomplete}.
For some of these models, we also have an orbifold
construction~\cite{GP}. Clarke recovered both the Batyrev--Borisov
and Hori--Vafa constructions from a duality for toric LG
models~\cite{Clarke}. The relation between the sphere
partition functions of mirror GLSMs was studied in~\cite{MMP}.
What we would like to understand here is how the two
GLSMs can be obtained from the same parent theory.

The rest of this paper is organized as follows.
In Section~\ref{sec:original}, we review the
Batyrev--Borisov construction and the Morrison--Plesser
parent theory, with particular attention to the sign issue.
In Section~\ref{sec:HV}, we discuss this issue from the
point of view of the Hori--Vafa mirror construction.
We then introduce our extended parent theory in
Section~\ref{sec:Higgs} and show in
Section~\ref{sec:reductions} how the sign issue can be
resolved.
In Section~\ref{sec:fixed}, we discuss the IR physics
of the resulting superpotential at finite $L$.
Further evidence for our proposal is given in
Section~\ref{sec:conjecture}, where we also discuss
the limitations of our argument for the equivalence
of the two IR SCFTs.
Two examples, the quintic and the complete intersection
of two cubics in $\mathbb P^5$, are discussed in
Section~\ref{sec:examples}.
We conclude in Section~\ref{sec:discussion} with some comments on the remaining problems and possible future directions. Some details of the calculations are given
in the appendices.

\section{Batyrev--Borisov data and the Morrison--Plesser obstruction}
\label{sec:original}

Before discussing the Morrison--Plesser construction, let us
first explain our notation. We use $\Phi_i$ and
$\widehat\Phi_a$ to denote the matter superfields in the MP
parent theory, and $\Sigma_H$ and $\widehat\Sigma$ to denote
the corresponding field-strength superfields. Later, in our
extended parent theory, we will introduce additional matter
superfields $\Theta,\Xi,S,T,N$ and field strengths $\Sigma_K$.

In this paper, we distinguish a superfield from its lowest
component whenever necessary. In particular, we use the
following notation:
\begin{equation}
 w=W|_0,\qquad \widetilde w=\widetilde W|_0,\qquad
 \phi_i=\Phi_i|_0,\quad \widehat\phi_a=\widehat\Phi_a|_0,
 \qquad b_i=B_i|_0.
 \label{eq:superfieldnotation}
\end{equation}
Thus, when we discuss the critical locus or compute the
Hessian, we take derivatives of $w$ or $\widetilde w$ with
respect to the corresponding scalar fields. Note that the
scalar potential in the Lagrangian also depends on the
kinetic metric. Similarly, we use $p_\alpha$ and $u_j$ for
the lowest components of the superfields $\mathcal P_\alpha$
and $\mathcal U_j$, respectively, and the same notation applies
to the hatted fields. Since the hatted matter fields in the
parent theory are twisted chiral superfields, both $W_L$ and
$W_{\rm BB}$ appearing later in this paper are twisted
superpotentials before the mirror exchange.

\subsection{Batyrev--Borisov data and the exponent matrix}
\label{subsec:BBdata}

Let us first recall how the two GLSMs associated with a
Batyrev--Borisov mirror pair can be described in terms of
an integer matrix.

Consider a $D$-dimensional Gorenstein toric Fano variety $V$, with
corresponding dual lattices $M$ and $N$. Let $v_j\in N$ be the primitive
generators of the rays of the fan of $V$, before any crepant refinement,
and $D_j^{\rm tor}$ the corresponding divisors. Recall that
$-K_V=\sum_j D_j^{\rm tor}$. Divide the rays into $r$ sets
$I_1,\ldots,I_r$ and define
\[
 D_\alpha=\sum_{j\in I_\alpha}D_j^{\rm tor},\qquad
 \sum_{\alpha=1}^r D_\alpha=-K_V.
\]
If all $D_\alpha$ are nef Cartier divisors, we have a nef partition~\cite[Section 4]{BBcomplete}. We are interested in the complete intersection
\[
 Y=\{G_1=\cdots=G_r=0\}\subset V,\qquad
 G_\alpha\in H^0(V,\mathcal O_V(D_\alpha)).
\]
For a smooth transverse complete intersection, adjunction gives
$K_Y=(K_V+\sum_\alpha D_\alpha)|_Y\simeq\mathcal O_Y$,
and $\dim_{\mathbb C}Y=D-r$.

In the homogeneous coordinates $\mathcal U_j$ of $V$~\cite{Cox}, the
polynomial $G_\alpha$ is a linear combination of monomials labelled by
$m\in\Delta_\alpha\cap M$, where
\begin{equation}
 \Delta_\alpha=\{m\in M_{\mathbb R}:
  \langle m,v_j\rangle\geq-\mathbf 1_{j\in I_\alpha}
  \text{ for every }j\}.
 \label{eq:BBNewtonPolytope}
\end{equation}
The reason for the above inequalities can be seen from the monomial
associated with $m$,
\begin{equation}
 M_{\alpha,m}(\mathcal U)
 =\prod_j\mathcal U_j^{\langle m,v_j\rangle+
                         \mathbf 1_{j\in I_\alpha}}.
 \label{eq:BBCoxMonomial}
\end{equation}
Indeed, the character associated with $m$ has divisor
$\sum_j\langle m,v_j\rangle D_j^{\rm tor}$. Adding $D_\alpha$ gives the
exponents in the above formula, which should all be nonnegative for
a section of $\mathcal O_V(D_\alpha)$. Each $\Delta_\alpha$ contains the
origin, and $\Delta=\sum_\alpha\Delta_\alpha$ is reflexive.

For the mirror construction, Borisov associates to each set $I_\beta$
a polytope in $N_{\mathbb R}$,
\begin{equation}
 \nabla_\beta=\operatorname{Conv}\bigl(\{0\}\cup
                         \{v_j:j\in I_\beta\}\bigr).
 \label{eq:BBDualNef}
\end{equation}
From the definition of $\Delta_\alpha$, we have
\begin{equation}
 \langle\Delta_\alpha,\nabla_\beta\rangle
 \geq-\delta_{\alpha\beta}.
 \label{eq:BBpairing}
\end{equation}
An important result of the BB construction is that
$\nabla=\sum_\beta\nabla_\beta$ is also reflexive, and
\begin{equation}
 \Delta^\circ=\operatorname{Conv}\Bigl(\bigcup_\beta\nabla_\beta\Bigr),
 \qquad
 \nabla^\circ=\operatorname{Conv}\Bigl(\bigcup_\alpha\Delta_\alpha\Bigr),
 \label{eq:BBPolarRelations}
\end{equation}
where $\circ$ denotes the polar polytope in the convention
$\langle m,n\rangle\geq-1$~\cite[Section 4]{BBcomplete}. One can equivalently
obtain $\nabla_\beta$ from the inequalities
\[
 \nabla_\beta=\{n\in N_{\mathbb R}:\langle m,n\rangle\geq
 -\delta_{\alpha\beta}\text{ for all }m\in\Delta_\alpha
 \text{ and all }\alpha\}.
\]
The mirror Calabi--Yau is defined by $r$ equations with Newton
polytopes $\nabla_\beta$ in the toric variety associated with $\nabla$
(with the normal fan of $\nabla$), possibly after a crepant refinement.
For a hypersurface ($r=1$), we recover Batyrev's construction with
$\nabla=\Delta^\circ$.

For example, for the complete intersection of two cubics in $\mathbb P^5$,
we have $v_j=e_j$ for $j=1,\ldots,5$ and $v_6=-\sum_{j=1}^5e_j$.
One can take $I_1=\{1,2,3\}$ and $I_2=\{4,5,6\}$, so both $D_1$
and $D_2$ are of degree three. The two $\nabla$ polytopes are given by
\eqref{eq:BBDualNef}. Taking $m=(2,-1,-1,0,0)\in\Delta_1$ in
\eqref{eq:BBCoxMonomial}, for instance, gives $\mathcal U_1^3$.
We will discuss the GLSMs of this example in Section~\ref{subsec:twocubics}.

Let us now describe the BB data in a way that is more convenient for
the GLSM. To write the superpotential $\sum_\alpha\mathcal P_\alpha
G_\alpha$, we introduce $r$ extra chiral superfields $\mathcal P_\alpha$.
Without the superpotential, the target in the geometric phase is the
total space
\[
 X=\operatorname{Tot}\left(
 \bigoplus_{\alpha=1}^r\mathcal O_V(-D_\alpha)\right).
\]
Thus, we have a toric space of dimension $D+r$ rather than $D$. We
introduce the lattices
\[
    \overline M=M\oplus\mathbb Z^r,\qquad
    \overline N=N\oplus\mathbb Z^r.
\]
In $\overline N$, the rays for $\mathcal U_j$ and $\mathcal P_\beta$
are $(v_j,e_{\beta(j)})$ with $j\in I_{\beta(j)}$, and $(0,e_\beta)$,
respectively. Here the $e_\beta$ in the second factor are the standard
basis of $\mathbb Z^r$. These rays generate the support cone of the
fan of $X$.
A lattice point $m\in\Delta_\alpha$ is then represented by
$(m,e_\alpha)$ in $\overline M$, where $e_\alpha$ is the
$\alpha$-th basis vector of $\mathbb Z^r$. We do the same
for the lattice points on the mirror side. In this way,
we obtain the two dual Cayley cones
\begin{equation}
 K=\operatorname{Cone}\{(m,e_\alpha):m\in\Delta_\alpha\},
 \qquad
 K^\vee=\operatorname{Cone}\{(n,e_\beta):n\in\nabla_\beta\}.
 \label{eq:BBcones}
\end{equation}
Both cones have dimension $\rho=D+r$ and Gorenstein
index $r$. The corresponding degree elements are
\[
 \deg=(0,1,\ldots,1)\in\overline M,\qquad
 \deg^\vee=(0,1,\ldots,1)\in\overline N,
\]
which satisfy
$\langle\deg,\deg^\vee\rangle=r$~\cite{Borisov,BB}.

We denote the degree-one lattice points in $K$ by
\[
    \bar m_a=(m_a,e_{\alpha(a)}),\qquad a=1,\ldots,u,
\]
and those in $K^\vee$ by
\[
    \bar n_i=(n_i,e_{\beta(i)}),\qquad i=1,\ldots,n.
\]
For the full toric supports, we include all
these lattice points. In particular, on each side there
are $r$ points of the form $(0,e_\alpha)$, which will
play a special role in the GLSM description.

Let $\mathsf S$ be the matrix whose rows are given by $\bar m_a$, and
$\mathsf T$ be the matrix whose columns are given by $\bar n_i$.
By pairing the lattice points on the two sides, we obtain
the integer matrix
\begin{equation}
 P_{ai}=\langle \bar m_a,\bar n_i\rangle
       =\langle m_a,n_i\rangle
        +\delta_{\alpha(a),\beta(i)},\qquad
 P=\mathsf S\mathsf T,\qquad \operatorname{rank}P=\rho.
 \label{eq:BBP}
\end{equation}
The rank of $P$ follows from the fact that the full sets
of lattice points span the corresponding cones, so that
both $\mathsf S$ and $\mathsf T$ have rank $\rho$. Note that $P$ is not
necessarily a square matrix. One can also consider a
subfamily by keeping only some of the lattice points.
In that case, one needs to check the resulting cones,
the rank of $P$, and the critical locus for generic
coefficients separately.

In the GLSM before a toric refinement, the columns of $\mathsf T$ are given
by the lifted rays for $\mathcal U_j$ and $\mathcal P_\beta$ above.
In particular, for $\bar m_a=(m_a,e_{\alpha(a)})$,
\begin{align}
 \langle \bar m_a,(v_j,e_{\beta(j)})\rangle
 &=\langle m_a,v_j\rangle+\mathbf 1_{j\in I_{\alpha(a)}},
 \nonumber\\
 \langle \bar m_a,(0,e_\beta)\rangle&=\delta_{\alpha(a),\beta}.
\end{align}
These are precisely the powers of the fields in
$\mathcal P_{\alpha(a)}M_{\alpha(a),m_a}$. The same statement holds on
the mirror side. If we include more degree-one lattice points as
columns or rows, we obtain a larger GLSM presentation, and we should
choose a compatible fan (or FI chamber). Not all of these points are
rays of the original fan of $V$.

Consider an ambient GLSM charge vector $\mathsf Q^A$ satisfying
$\sum_j\mathsf Q_j^A v_j=0$. In the extended lattice, we have the
relation
\begin{equation}
 \sum_j\mathsf Q_j^A(v_j,e_{\beta(j)})
 -\sum_{\beta=1}^r\mathsf D_\beta^A(0,e_\beta)=0,
 \qquad
 \mathsf D_\beta^A=\sum_{j\in I_\beta}\mathsf Q_j^A.
 \label{eq:BBLiftedCharges}
\end{equation}
In other words, $-\mathsf D_\beta^A$ is the charge of
$\mathcal P_\beta$, which is precisely the charge needed for the
superpotential $\mathcal P_\beta G_\beta$. If we work with a refined
presentation, we use the relations among its rays instead.

Let us now see how to obtain the two GLSMs from $P$.
From \eqref{eq:BBpairing}, all the entries of $P$ are
nonnegative integers. We can therefore associate each
row of $P$ with a monomial in the ordinary chiral
superfields $\Phi_i$. Similarly, each column of $P$
gives a monomial in the twisted chiral superfields
$\widehat\Phi_a$ on the mirror side:
\begin{equation}
 A_a=\prod_i\Phi_i^{P_{ai}},\qquad
 B_i=\prod_a\widehat\Phi_a^{P_{ai}}.
 \label{eq:BBmonomials}
\end{equation}
In this regard, a field in one GLSM corresponds to
a monomial in the other GLSM.

Recall that we have included the lattice points
$(0,e_\beta)$ in $K^\vee$. Let us consider the column
of $P$ associated with such a point. From
\eqref{eq:BBP}, we have
\[
    \langle \bar m_a,(0,e_\beta)\rangle
    =\delta_{\alpha(a),\beta}.
\]
We denote the $r$ superfields associated with these
columns by $\mathcal P_\beta$. It is clear that each
monomial $A_a$ contains only one of these superfields,
namely $\mathcal P_{\alpha(a)}$, and that it is linear
in this field. Similarly, the lattice points $(0,e_\alpha)$
in $K$ give the $r$ twisted chiral superfields
$\widehat{\mathcal P}_\alpha$ on the mirror side.

Thus, if we denote the remaining matter superfields
on the two sides by $\mathcal U$ and $\widehat{\mathcal U}$,
respectively, we can write the monomials and the
superpotentials as follows:
\begin{align}
 A_a&=\mathcal P_{\alpha(a)}M_a(\mathcal U),\qquad
 B_i=\widehat{\mathcal P}_{\beta(i)}
       \widehat M_i(\widehat{\mathcal U}),
 \nonumber\\
 W_A&=\sum_\alpha\mathcal P_\alpha G_\alpha(\mathcal U),
 \qquad
 W_{\rm BB}=\sum_\beta\widehat{\mathcal P}_\beta
                  \widehat G_\beta(\widehat{\mathcal U}),
 \label{eq:BBsplit}
\end{align}
where
\[
 G_\alpha=\sum_{a:\alpha(a)=\alpha}c_aM_a,\qquad
 \widehat G_\beta
 =\sum_{i:\beta(i)=\beta}d_i\widehat M_i.
\]
These are the superpotentials we expect for the GLSMs
describing the two Calabi--Yau complete intersections.
Note that we did not assume the existence of a
Landau--Ginzburg phase with an isolated critical point.

One can also see from $P$ that the two theories have
the same central charge, provided that they flow to
compact SCFTs with the expected superconformal
R-symmetries. For this purpose, we introduce two
vectors $\boldsymbol\mu$ and $\boldsymbol\nu$. We set $\mu_i=1$ for the
$\mathcal P$ fields and $\mu_i=0$ for the other ordinary fields.
Similarly, we set $\nu_a=1$ for the $\widehat{\mathcal P}$ fields and zero
for the other fields. From the above discussion,
we have
\begin{equation}
 P\boldsymbol\mu=\mathbf1_u,\qquad P^T\boldsymbol\nu=\mathbf1_n,\qquad
 \sum_i\mu_i=\sum_a\nu_a=r.
 \label{eq:BBgrading}
\end{equation}
In the geometric phase, we can assign
$R(\Phi_i)=2\mu_i$ and $R(\widehat\Phi_a)=2\nu_a$,
where we use the corresponding mirror R-charge
convention for the hatted fields.

Recall that there are $n$ matter fields in the original
GLSM, while the rank of the gauge group is $n-\rho$.
The GLSM R-anomaly calculation of~\cite[Section 2]{MP} then gives
\[
 \widehat c_A
 =\sum_i(1-2\mu_i)-(n-\rho)
 =\rho-2\sum_i\mu_i.
\]
Doing the same calculation on the mirror side, we obtain
\begin{equation}
 \widehat c_A=\rho-2\sum_i\mu_i
             =\rho-2r
             =\rho-2\sum_a\nu_a=\widehat c_B,
 \qquad
 \dim_{\mathbb C}Y=D-r=\rho-2r.
 \label{eq:BBdimension}
\end{equation}
Here $\widehat c=c/3$ when the R-current used in the
calculation is the superconformal R-current.
Of course, this calculation concerns the two GLSMs
rather than the full graded parent theory. The
identification of the R-current in the latter theory
is a separate question.

\subsection{Integral gauge data and coefficient coordinates}
\label{subsec:BBintegral}

So far, we have discussed how to obtain the superpotentials
of the two GLSMs from the matrix $P$. Let us now discuss
their gauge groups. As explained in~\cite{MP}, the integer
factorization $P=\mathsf S\mathsf T$ determines the gauge groups through
\begin{equation}
 G=\ker\bigl(\mathsf T:U(1)^n\longrightarrow U(1)^\rho\bigr),
 \qquad
 \widehat G
 =\ker\bigl(\mathsf S^T:U(1)^u\longrightarrow U(1)^\rho\bigr).
 \label{eq:BBgroups}
\end{equation}
To understand this, recall that the charge vector
$\ell$ of a $U(1)$ gauge symmetry in the original GLSM
must satisfy $P\ell=0$, since otherwise some monomials
in the superpotential would not be gauge invariant.
Similarly, the charge vectors on the mirror side
satisfy $P^T\widehat\ell=0$. Thus the charge lattices
of the connected gauge groups are
$\ker_{\mathbb Z}P$ and $\ker_{\mathbb Z}P^T$,
respectively.

However, these charge lattices do not tell us whether
there are additional finite gauge symmetries. To obtain
this information, we need to know the integer maps
$\mathsf T$ and $\mathsf S^T$ in \eqref{eq:BBgroups}. In this formulation,
the mirror construction exchanges
\[
    (P,\mathsf S,\mathsf T)\quad\longleftrightarrow\quad
    (P^T,\mathsf T^T,\mathsf S^T).
\]
We also need to choose the FI chambers of the two GLSMs
to specify their phases.

Note that only the matrix $P$ appears explicitly in
the local interactions of the MP parent theory.
One may therefore ask how the information about the
two gauge groups, in particular their finite parts,
can be recovered from the parent theory. To answer
this question, one needs to study the full gauge
integral, including the sum over gauge bundles.
We will come back to this issue when discussing
our infrared conjecture.

Let us next consider the parameters in the two GLSMs.
Suppose that the coefficients of the monomials $B_i$
and $A_a$ are $v_i$ and $c_a$, respectively, and that
all these coefficients are nonzero. For any
$\ell\in\ker_{\mathbb Z}P$ and
$\widehat\ell\in\ker_{\mathbb Z}P^T$, we can form
the following combinations:
\begin{equation}
 z_\ell(v)=\prod_i v_i^{\ell_i},\qquad
 \widehat z_{\widehat\ell}(c)
 =\prod_a c_a^{\widehat\ell_a}.
 \label{eq:BBcharacters}
\end{equation}
The reason for considering these combinations is
that the individual coefficients can be changed by
rescaling the matter fields. For example, consider
a rescaling of the hatted field $\widehat\Phi_a$.
The coefficient of $B_i$ then transforms with a
power determined by $P_{ai}$. Since $P\ell=0$,
these factors cancel in $z_\ell(v)$. Similarly,
$\widehat z_{\widehat\ell}(c)$ is invariant under
rescalings of the ordinary matter fields.

Contracting $P^T\boldsymbol\nu=\mathbf1_n$ with $\ell$ and
$P\boldsymbol\mu=\mathbf1_u$ with $\widehat\ell$ gives
\[
    \sum_i\ell_i=\boldsymbol\nu^TP\ell=0,\qquad
    \sum_a\widehat\ell_a=\widehat\ell^TP\boldsymbol\mu=0.
\]
We will use this property later when comparing the
FI parameters and the superpotential coefficients
in our extended parent theory.

Note that in \eqref{eq:BBcharacters} we have only
taken into account the diagonal field rescalings.
More general field redefinitions, including
redefinitions of the equations defining the complete
intersection, can further reduce the number of
independent parameters. Also, some of the operators
in the GLSM description may become trivial in the IR.
On the other hand, not all complex-structure or
K\"ahler deformations of a Calabi--Yau manifold can
necessarily be realized in terms of these GLSM
parameters.

The local calculation that we will discuss below
can be applied to each column of $P$, independently
of which phase we eventually consider. However,
when discussing the IR physics, we need to take
the phase of the GLSM into account. For example,
in the geometric phase, we will use the transversality
of the equations defining a smooth complete
intersection. Note that a crepant partial resolution
does not necessarily remove all the singularities
in higher-dimensional cases~\cite{BBcomplete}.
In the quintic example, one can also study the IR
physics from its LG phase, where the superpotential
has an isolated critical point. In a hybrid phase,
one should keep the fields that remain massless
along the critical locus.

\subsection{The doubled action and logarithmic signs}

A difficulty in applying abelian duality to the GLSM is that, in
general, the phase rotations $H=U(1)^n$ of the chiral fields are not
symmetries once we turn on $W_A$. Morrison and Plesser's idea was to
replace this breaking of the symmetry by an anomaly. They introduced
$u$ twisted chiral superfields, each coupled to its own twisted gauge
multiplet, and replaced the polynomial superpotential by
\[
 W_s=C\sum_a\widehat\Sigma_a[\log(c_aA_a/\mu)+1],
 \qquad C=\frac{1}{2\pi\sqrt2}.
\]
Under $\Phi_i\mapsto e^{i\theta_i}\Phi_i$, we have
\[
 \delta W_s=iC\sum_a\widehat\Sigma_a(P\theta)_a.
\]
The change is now a shift of the theta angles of the new gauge fields.
We will see below how the polynomial $W_A$ is recovered by integrating
out the new massive matter fields and then $\widehat\Sigma_a$.

In the usual abelian duality, if one gauges all of $H$, one needs
additional superfields imposing the constraint that the new field
strengths lie in $\mathfrak g=\operatorname{Lie}(G)$. In the present
model these constraint superfields would themselves be charged under
the twisted gauge group. MP pointed out that their
one-loop contribution need not go away even when their kinetic term
is removed. Instead of this procedure, they considered a theory
with both sets of gauge and matter multiplets~\cite[Section 5]{MP}.

In this parent theory, each
ordinary matter field $\Phi_i$ is coupled to a
$U(1)_{H_i}$ gauge field, and each twisted matter
field $\widehat\Phi_a$ is coupled to a
$U(1)_{\widehat H_a}$ gauge field. We denote the
corresponding field-strength superfields by
$\Sigma_i\equiv\Sigma_{H_i}$ and $\widehat\Sigma_a$,
which are twisted chiral and chiral superfields,
respectively.

The interactions in the MP parent theory are given
in eqs.~(14)--(15) of~\cite{MP}. In our notation,
they read
\begin{align}
 K_{\rm mix}&=\frac{2}{\pi}V^TP^T\widehat V,
 \nonumber\\
 W_{\rm MP}&=C\sum_a\widehat\Sigma_a
       [\log(c_a A_a/\mu)+1],
 \nonumber\\
 \widetilde W_{\rm MP}&=C\sum_i\Sigma_i
       [\log(a_i/(\mu B_i))+1],
 \qquad C=\frac{1}{2\pi\sqrt2}.
 \label{eq:MP}
\end{align}
Here $a_i$ and $c_a$ are the holomorphic FI
coefficients of mass dimension one, and $\mu$
is a reference scale.

An important observation is that $A_a$ and $B_i$
appear with opposite powers inside the logarithms.
As we will see below, this is precisely the reason
why the two integration processes in the MP
parent theory do not give the expected mirror pair.
Before discussing the one-loop calculation, let
us first understand where this relative sign comes
from.

Recall that Morrison and Plesser originally
introduced a gauge-invariant D-term interaction,
given in eq.~(13) of~\cite{MP},
\begin{align}
 K_{\rm MP}^{(13)}
 &=\sum_i\left(J_i-\frac{|\Sigma_i|^2}{4e_i^2}\right)
 -\sum_a\left(\widehat J_a
             -\frac{|\widehat\Sigma_a|^2}
                    {4\widehat e_a^2}\right)
 \nonumber\\
 &\quad+\frac1{2\pi}\sum_{ai}P_{ai}
             \log J_i\log\widehat J_a ,
 \label{eq:MP13}
\end{align}
where $J_i=\overline\Phi_i e^{2V_i}\Phi_i$, and
$\widehat J_a$ is defined similarly in terms of
the hatted fields.

Locally, where the logarithms are well defined,
one can expand the last term in \eqref{eq:MP13}.
By integration by parts in superspace, the term
containing $\log\Phi_i\,\widehat V_a$ gives a
coupling between $\log\Phi_i$ and $\widehat\Sigma_a$.
Similarly, the term containing
$V_i\log\widehat\Phi_a$ gives a coupling between
$\Sigma_i$ and $\log\widehat\Phi_a$. The relative
sign of these two couplings is fixed by the
requirement that their gauge variations cancel
the gauge variation of the remaining
$V\widehat V$ term, up to a total derivative.
MP distinguish gauge transformations trivial at infinity
from the constant phase rotations, which are still anomalous.

Note that in the presence of instantons, the
integration by parts may give additional boundary
terms. Morrison and Plesser therefore use
\eqref{eq:MP}, together with the kinetic terms,
to define their quantum theory.

There is another point which will be useful for
our discussion of the integration measure.
From \eqref{eq:MP}, we see that $\log\Phi_i$
is coupled to a dynamical field
\[
    U_i=C(P^T\widehat\Sigma)_i.
\]
For any $\ell\in\ker_{\mathbb Z}P$, we have
\[
    \ell^TU=0.
\]
Since the $\widehat\Sigma_a$ are dynamical fields
rather than external sources, these logarithmic
couplings remain in the theory even when we turn
off the external sources. The same is true for
the mixed D-term interaction.

\subsection{The critical-point equation and inverse powers}

We are now ready to discuss the two integration
processes in the MP parent theory. Let us first
consider a general one-loop calculation which
will be used on both sides.

Suppose that we have a unit-charge matter
multiplet whose mass is determined by a
field-strength superfield $\mathcal S$.
Integrating out this matter multiplet gives
a one-loop contribution of the form~\cite[eq.~(3.39)]{HV}
\[
    -\mathcal S
       [\log(\mathcal S/\mu)-1].
\]
If the tree-level superpotential also contains
a logarithmic coupling
$\mathcal S\log(\mathcal M/\mu)$, the resulting
effective superpotential, up to an overall
normalization, is
\begin{equation}
 W_{\rm stat}(\mathcal S;\mathcal M)
 =\mathcal S\log(\mathcal M/\mu)
  -\mathcal S[\log(\mathcal S/\mu)-1].
 \label{eq:MPstationary}
\end{equation}
Here we have absorbed the linear counterterm
into the FI coefficient. Depending on the
chirality of $\mathcal S$, this is either an
ordinary or a twisted superpotential.

Let $z=\mathcal S|_0$ and $m=\mathcal M|_0$.
To integrate out $\mathcal S$, we solve the
critical-point equation
\[
    \frac{\partial w_{\rm stat}}{\partial z}
    =\log\left(\frac{m}{z}\right)=0,
\]
which gives $z_*=m$. Substituting this solution
back into the effective superpotential, we
obtain
\[
    w_{\rm stat}|_*=m.
\]

Let us first apply this calculation to the
integration process in which we recover the
original GLSM. In a region where the lowest components of
$\widehat\Sigma_a$ are nonzero and the hatted matter fields
are massive, we can integrate out these matter fields. From \eqref{eq:MP},
we identify
\[
    \mathcal S=\widehat\Sigma_a,\qquad
    \mathcal M=c_aA_a.
\]
As in eq.~(12) of~\cite{MP}, eliminating $\widehat\Sigma_a$ gives the
superpotential term $c_aA_a$ for each row
of $P$. Summing over all the rows, we obtain
the expected superpotential of the original
GLSM.

One may expect that the same calculation,
with the roles of the ordinary and twisted
chiral fields exchanged, would give the
mirror superpotential. Unfortunately, this
is not the case. If we keep the hatted
matter fields fixed and integrate out the
ordinary matter fields, we instead have
\[
    \mathcal S=\Sigma_i,\qquad
    \mathcal M=\frac{a_i}{B_i}.
\]
Thus, after solving the critical-point
equation, we obtain
\begin{equation}
 \Sigma_{i,*}=\frac{a_i}{B_i},\qquad
 \widetilde W_-=\sum_i\frac{a_i}{B_i}.
 \label{eq:MPinverse}
\end{equation}
In the component description, the first
equation reads $\sigma_{i,*}=a_i/b_i$,
where $\sigma_i=\Sigma_i|_0$.

Note that the one-loop contributions have
the same sign in the two integration
processes. The difference comes from the
tree-level logarithmic couplings in
\eqref{eq:MP}, where $A_a$ appears in the numerator while $B_i$ appears in the denominator. As discussed above, we cannot simply change the sign of the logarithmic coupling on the mirror side, since doing
so would spoil the gauge invariance of the parent theory. This is the sign issue encountered by Morrison and Plesser~\cite{MP}.

\subsection{The gauge symmetries in the two reductions}
\label{subsec:MPgaugeReduction}

To recover the original GLSM, one needs not only $W_A$ but also the
gauge group $G$. Let $Q^A$ be a basis of $\ker_{\mathbb Z}P$. We can
locally write the ordinary vector multiplets as
\[
 V=\sum_A Q^A V_A+V_\perp,
\]
where $V_\perp$ denotes the vector multiplets in the complementary
directions. Since $PQ^A=0$, we have
\[
 K_{\rm mix}=\frac{2}{\pi}(PV_\perp)^T\widehat V.
\]
Also, the field-dependent term in $\widetilde W_{\rm MP}$ is
\begin{equation}
 -C\sum_i\Sigma_i\log B_i
 =-C\sum_a(P\Sigma)_a\log\widehat\Phi_a.
 \label{eq:MPGaugeDirections}
\end{equation}
Thus, the $V_A$ do not enter these mixed interactions, and all $A_a$
are invariant under the corresponding gauge transformations. The
remaining ordinary gauge multiplets do couple to the hatted system.
If the latter can be integrated out with the constraint $P\Sigma=0$
on the surviving gauge fields, we obtain the gauge Lie algebra of
$G$. The analogous equation on the B side is
$P^T\widehat\Sigma=0$. MP argued for such a reduction of the gauge
group~\cite[Section 5]{MP}, but the above one-loop calculation of
the superpotential does not itself establish these constraints.

In fact, even if the continuous gauge symmetry has been recovered,
we still need to determine the finite gauge group. Regard $\mathsf S$ and
$\mathsf T$ as maps between compact tori. Since $\mathsf T$ is surjective and $\mathsf S$
has finite kernel, the factorization $P=\mathsf S\mathsf T$ gives
\begin{equation}
 1\longrightarrow\ker \mathsf T\longrightarrow\ker P
 \xrightarrow{\ \mathsf T\ }\ker \mathsf S\longrightarrow1.
 \label{eq:MPFiniteKernel}
\end{equation}
On the mirror side, we have
\[
 1\longrightarrow\ker \mathsf S^T\longrightarrow\ker P^T
 \xrightarrow{\ \mathsf S^T\ }\ker \mathsf T^T\longrightarrow1.
\]
For example, $G=\ker \mathsf T$ need not be the entire group $\ker P$
which preserves $W_A$, although they have the same Lie algebra.
This is why the integral factorization of $P$ is important. The
local parent action only contains $P$, so one has to specify the
global gauge data and the corresponding gauge bundle sums as well.

\subsection{Parameter matching and redundant deformations}
\label{subsec:MPredundant}

Assuming the gauge reduction above, we can read off the FI parameters
by substituting $\Sigma_i=\sum_AQ_i^A\Sigma_A$ into the ordinary FI term.
For $\ell\in\ker_{\mathbb Z}P$, this gives the algebraic FI coordinate
\begin{equation}
 q_\ell=\prod_i(a_i/\mu)^{\ell_i}=\prod_i a_i^{\ell_i}.
 \label{eq:MPordinaryFI}
\end{equation}
Here we have used $\sum_i\ell_i=0$, so the linear counterterm also
drops out. Similarly, $\widehat q_{\widehat\ell}=\prod_a c_a^{\widehat\ell_a}$.
If the B-side calculation had given $\sum_i a_i B_i$, we would have
precisely the parameter correspondence of~\cite[eq.~(9)]{MP}.

There is another important point in the MP argument. Not all the
parameters $a_i$ and $c_a$ are independent. Consider the field
redefinitions $\Phi_i'=e^{\lambda_i}\Phi_i$ and
$\widehat\Phi_a'=e^{\widehat\lambda_a}\widehat\Phi_a$.
Writing the logarithmic couplings in terms of the primed fields,
we obtain the new coefficients
\begin{equation}
 c_a'=c_a e^{-(P\lambda)_a},\qquad
 a_i'=a_i e^{(P^T\widehat\lambda)_i}.
 \label{eq:MPrescalings}
\end{equation}
From $P\ell=0$ and $P^T\widehat\ell=0$, we have
\begin{align}
 \prod_i(a_i')^{\ell_i}&=\prod_i a_i^{\ell_i},
 &\prod_a(c_a')^{\widehat\ell_a}&=\prod_a c_a^{\widehat\ell_a}.
\end{align}
In particular, since $P\boldsymbol\mu=\mathbf1_u$ and $P^T\boldsymbol\nu=\mathbf1_n$, we
can take $\lambda=\gamma\boldsymbol\mu$ and
$\widehat\lambda=\widehat\gamma\boldsymbol\nu$, with
$\gamma,\widehat\gamma\in\mathbb C$, to rescale
all the $c_a$ or all the $a_i$ by a common factor.

The idea of MP is to use redundant deformations to reach two regions
of the parent theory: in one region the hatted fields can be integrated
out, and in the other region the ordinary fields can be integrated out.
If the two calculations give the two GLSMs without missing any massless
fields, and the deformations do not change the IR physics, then the
IR SCFTs of the two GLSMs are isomorphic under mirror symmetry.
Unfortunately, the B-side superpotential has the wrong powers.

Of course, a field redefinition also changes the kinetic terms, and
one has to account for the measure in the quantum theory. An exact
change of variables alone cannot change the physical masses. Hence,
keeping the algebraic FI parameters fixed is not sufficient to show
that varying the microscopic couplings at fixed kinetic terms is
redundant. We will come back to this issue for our extended parent
theory.

In the next section, we will first see whether the negative powers can be understood from the Hori--Vafa mirror construction.

\section{Mirror variables and the integration measure}
\label{sec:HV}

Before modifying the Morrison--Plesser parent theory, we
would like to see whether the sign issue discussed in the
previous section can be understood from the Hori--Vafa
mirror construction. An interesting observation in the
Hori--Vafa construction is that the mirrors of a compact
Calabi--Yau complete intersection and its associated
noncompact Calabi--Yau manifold can have the same formal
superpotential. The difference between the two theories
is reflected in the choice of fundamental variables and
the integration measure. We first review this observation
for general complete intersections and then discuss its
relation to the MP construction.

\subsection{Compact complete intersections and their noncompact models}

Consider a GLSM with gauge group $U(1)^k$ for its connected
part and matter superfields $\mathcal U_j$, $j=1,\ldots,N$,
of charges $\mathsf Q_j^A$. We are interested in a complete
intersection defined by $r$ equations, where the
$\lambda$-th equation is a section of a line bundle of
multidegree $\mathsf D_\lambda^A$. The Calabi--Yau condition
is
\begin{equation}
 \sum_j\mathsf Q_j^A=\sum_\lambda\mathsf D_\lambda^A.
 \label{eq:HVCYgeneral}
\end{equation}
To describe the complete intersection, we introduce
$r$ chiral superfields $\mathcal P_\lambda$ of charges
$-\mathsf D_\lambda^A$ and turn on the superpotential
\[
 W_A=\sum_{\lambda=1}^r
       \mathcal P_\lambda G_\lambda(\mathcal U).
\]

Let us first consider the theory without this
superpotential. In the geometric phase, the scalar fields
$p_\lambda=\mathcal P_\lambda|_0$ are now the fiber
coordinates of the dual line bundles over the toric
variety parametrized by $u_j=\mathcal U_j|_0$.
Thus, instead of a compact complete intersection, we
have a noncompact Calabi--Yau manifold given by the total
space of the direct sum of these line bundles.

Following Hori and Vafa, we perform abelian duality on
the matter fields and take into account the contributions
of vortices. We obtain the periodic twisted chiral
superfields $\mathcal Y_j$ and
$\mathcal Y_{p_\lambda}$ with twisted superpotential
\begin{align}
 \widetilde W_{\rm HV}
 &=\sum_{A=1}^k\Sigma_A\left(
     \sum_j\mathsf Q_j^A\mathcal Y_j
     -\sum_\lambda\mathsf D_\lambda^A\mathcal Y_{p_\lambda}
     -t_A\right)
   +\sum_je^{-\mathcal Y_j}
   +\sum_\lambda e^{-\mathcal Y_{p_\lambda}}.
 \label{eq:HVgeneral}
\end{align}
Here $t_A$ are the complexified FI parameters. We have set
the common scale of the exponential terms to one. As usual, when computing the periods, we
take the lowest components of the superfields, which
we denote by $y_j$, $y_{p_\lambda}$ and $\sigma_A$.
The integrand of the period contains
$e^{-\widetilde w_{\rm HV}}$, where
$\widetilde w_{\rm HV}=\widetilde W_{\rm HV}|_0$.

Hori and Vafa observed that the period associated with
the compact Calabi--Yau manifold can be obtained from
that of the noncompact model by taking derivatives
with respect to the FI parameters~\cite[eqs.~(7.75)--(7.78)]{HV}.
To see this, recall that taking a derivative with
respect to $t_A$ brings down a factor of $\sigma_A$
in the period integral. For the $\lambda$-th defining
equation, we consider the following combination of
derivatives:
\[
 \sum_{A=1}^k\mathsf D_\lambda^A
       \frac{\partial}{\partial t_A}.
\]
The corresponding insertion is
$\delta_\lambda=\sum_A\mathsf D_\lambda^A\sigma_A$.
For periods over corresponding cycles transported locally in FI
parameter space, the $r$ defining equations give
\begin{align}
 \Pi_{\rm c}(t)&\propto
  \left[\prod_{\lambda=1}^r
   \left(\sum_{A=1}^k\mathsf D_\lambda^A
          \frac{\partial}{\partial t_A}\right)\right]
          \Pi_{\rm nc}(t).
 \label{eq:HVgeneralderivative}
\end{align}

Let us see how these derivatives affect the measure
of the period integral. By taking the derivative of
$e^{-\widetilde w_{\rm HV}}$ with respect to
$y_{p_\lambda}$, we have
\[
 \frac{\partial}{\partial y_{p_\lambda}}
 e^{-\widetilde w_{\rm HV}}
 =
 \left(
 \sum_A\mathsf D_\lambda^A\sigma_A
 +e^{-y_{p_\lambda}}
 \right)e^{-\widetilde w_{\rm HV}}.
\]
On cycles for which the boundary term vanishes, integration
by parts lets us replace the insertion
$\sum_A\mathsf D_\lambda^A\sigma_A$ by
$-e^{-y_{p_\lambda}}$ in the integral. After integrating
out the $\sigma_A$ fields, we obtain
\begin{align}
 \Pi_{\rm c}(t)&\propto
  \int \prod_jdy_j
       \prod_\lambda\left(e^{-y_{p_\lambda}}
                          dy_{p_\lambda}\right)
       e^{-\sum_je^{-y_j}
          -\sum_\lambda e^{-y_{p_\lambda}}}
       \nonumber\\
 &\qquad\times\prod_A\delta\!\left(
          \sum_j\mathsf Q_j^A y_j
          -\sum_\lambda\mathsf D_\lambda^A
                              y_{p_\lambda}-t_A\right).
 \label{eq:HVgeneralmeasure}
\end{align}
Here we have absorbed the overall sign into the
normalization of the period. Note that this calculation
does not require the degree matrix
$\mathsf D_\lambda^A$ to be square or invertible.

There is another way to understand the additional
factors in the measure. In~\cite{GS}, the author and
Sharpe studied abelian duality in the presence of a
background R-symmetry gauge field. In the localized
zero-mode integral, a matter field of R-charge $R$
gives a measure of the form
\[
    e^{-Ry/2}dy.
\]
Recall that in the GLSM for a complete intersection,
we can assign R-charge zero to the $\mathcal U_j$
fields and R-charge two to the $\mathcal P_\lambda$
fields. Thus, the measure for the former is $dy_j$,
while for the latter it is
\[
    e^{-y_{p_\lambda}}dy_{p_\lambda}.
\]
This is precisely the measure appearing in
\eqref{eq:HVgeneralmeasure}. In particular, if we
introduce the variable $V_\lambda=e^{-y_{p_\lambda}}$,
the measure for the dual of $\mathcal P_\lambda$ is
proportional to $dV_\lambda$. The measure for the duals
of $\mathcal U_j$ is logarithmic in $e^{-y_j}$.

\subsection{The nef partition in the mirror period}

Let us now see how the nef partition appears in the
above mirror construction. We denote the partition
of the toric coordinates by
$E_1\sqcup\cdots\sqcup E_r$, so that
\[
 \mathsf D_\lambda^A
 =\sum_{j\in E_\lambda}\mathsf Q_j^A.
\]
For the coordinates belonging to $E_\lambda$, we
introduce the following variables on the dense torus:
\begin{equation}
 V_\lambda=e^{-y_{p_\lambda}},\qquad
 \mathsf U_j=\frac{e^{-y_j}}{V_\lambda}
       \quad(j\in E_\lambda).
 \label{eq:HVnefvariables}
\end{equation}
The terms $e^{-y_j}$ with $j\in E_\lambda$ and the term
$e^{-y_{p_\lambda}}$ now have a common factor $V_\lambda$.
In the $A$-th exponentiated gauge constraint its exponent is
$\sum_{j\in E_\lambda}\mathsf Q_j^A-\mathsf D_\lambda^A=0$.

In terms of these new variables, the constraints
and the holomorphic interaction become
\begin{equation}
 \prod_j\mathsf U_j^{\mathsf Q_j^A}=e^{-t_A},\qquad
 w_{\rm nef}=\sum_{\lambda=1}^r
              V_\lambda\left(1+\sum_{j\in E_\lambda}\mathsf U_j\right).
 \label{eq:HVnefpotential}
\end{equation}
Recall that for the compact period, the integration
measure for $V_\lambda$ is $dV_\lambda$ rather than
$dV_\lambda/V_\lambda$. Since $w_{\rm nef}$ is linear
in each $V_\lambda$, we can choose the corresponding
holomorphic integration cycles to express the
oscillatory integral as a residue integral. Integrating
over the $V_\lambda$ variables then gives the constraints
\begin{equation}
 1+\sum_{j\in E_\lambda}\mathsf U_j=0,
 \qquad \lambda=1,\ldots,r.
 \label{eq:HVnefconstraints}
\end{equation}
Together with the $k$ gauge constraints in
\eqref{eq:HVnefpotential}, these equations describe
a complete intersection of dimension $N-k-r$ in
the mirror torus. This is the mirror construction
discussed by Hori and Vafa in Section~7.3,
eqs.~(7.78)--(7.84) of~\cite{HV}. The compactification
of the mirror and the finite gauge identifications
are determined by the dual nef partition and the
corresponding integral lattice data.

Note that we did not need to assume the existence
of a polynomial LG phase in the above calculation.

As an example, let us consider the complete
intersection of two cubics in $\mathbb P^5$.
In this case, $k=1$ and $r=2$. We divide the six
toric coordinates into two groups of three.
From \eqref{eq:HVnefpotential} and
\eqref{eq:HVnefconstraints}, the mirror equations
are
\begin{equation}
 \prod_{j=1}^6\mathsf U_j=e^{-t},\qquad
 1+\mathsf U_1+\mathsf U_2+\mathsf U_3=0,\qquad
 1+\mathsf U_4+\mathsf U_5+\mathsf U_6=0.
 \label{eq:HVbicubic}
\end{equation}
Since both defining equations have degree three,
the period of the compact model is obtained by
acting with $(3\partial_t)^2$ on the period of the
noncompact model, up to an overall normalization.

For the quintic, there is only one defining
equation and hence only one derivative with respect
to the FI parameter. We will discuss this example
in more detail below, since it allows us to see
explicitly how the derivative changes the measure
in the period integral.

\subsection{Inverting the torus coordinates in the MP parent}

Let us now go back to the Morrison--Plesser
construction. Recall that the superpotential obtained
on the mirror side has negative powers of the
monomials $B_i$. One may ask whether the reason for
these negative powers is that we have not used the
right variables to describe the mirror theory.

For a general exponent matrix $P$, consider the
following change of variables on the dense torus:
\begin{equation}
 Z_a=\widehat\phi_a^{-1},\qquad
 \widetilde w_-=\sum_i\frac{a_i}{b_i(\widehat\phi)}
       =\sum_i a_i\prod_a Z_a^{P_{ai}}.
 \label{eq:MPgeneralinversion}
\end{equation}
In terms of the $Z_a$ variables, we have indeed
obtained the expected polynomial.

However, this does not immediately mean that we
have obtained the correct mirror theory. First,
the gauge charges of the new variables are opposite
to those of the original variables. Second, we have
only performed the field transformation on the
dense torus, where none of the fields vanishes.
When computing the periods, we also need to know
the integration cycle and the measure in the new
variables, as well as how the cycle extends across
the coordinate divisors.

From our discussion of the Hori--Vafa construction,
we expect that the measure for the compact model
contains one additional factor for each defining
equation of the complete intersection. Thus, if
integrating out the massive fields in the MP parent
theory could generate these factors, it might be
possible to understand the negative powers through
a change of variables without modifying the parent
theory.

We would like to test this possibility. For
simplicity, we first consider the quintic, where
there is only one such factor in the measure.

\subsection{The quintic measure test}
\label{subsec:quinticmeasure}

\subsubsection{Compact and noncompact periods}

Recall that the GLSM for the quintic and that for
the noncompact Calabi--Yau manifold
$\operatorname{Tot}(\mathcal O_{\mathbb P^4}(-5))$
have the same matter content and gauge charges.
The difference is that in the GLSM for the quintic
we turn on the superpotential
$\mathcal P G_5(\mathcal U)$.

Let us first consider the noncompact model.
The Hori--Vafa mirror superpotential is
\begin{equation}
 \widetilde W_{\rm HV}
 =\Sigma\left(\sum_{j=1}^5\mathcal Y_j-5\mathcal Y_p-t\right)
   +\sum_{j=1}^5e^{-\mathcal Y_j}+e^{-\mathcal Y_p}.
 \label{eq:HVdual}
\end{equation}
As before, we set the common scale of the exponential
terms to one.

By integrating out $\Sigma$, we have
\[
    5\mathcal Y_p=\sum_j\mathcal Y_j-t.
\]
We can solve this constraint by introducing
$\mathcal Z_j=e^{-\mathcal Y_j/5}$ on a fifth-root
cover. In particular, we have
\[
    e^{-\mathcal Y_p}=e^{t/5}\prod_j\mathcal Z_j.
\]
Thus, the mirror superpotential can be written as
\begin{equation}
 W_{\rm HV}(\mathcal Z)
 =\sum_j\mathcal Z_j^5+e^{t/5}\prod_j\mathcal Z_j,
 \label{eq:HVquintic}
\end{equation}
with identifications $\mathcal Z_j\mapsto\omega_j\mathcal Z_j$,
where $\omega_j^5=1$ and $\prod_j\omega_j=1$. The latter condition
keeps $e^{-\mathcal Y_p}$ fixed.

Although this is the familiar superpotential of
the LG mirror of the quintic, we should recall that
we started with the noncompact Calabi--Yau manifold.
The important difference can be seen from the
period integral.

Let $Z_j=\mathcal Z_j|_0$ and
$w_{\rm HV}(Z)=W_{\rm HV}(\mathcal Z)|_0$.
Since the fundamental integration variables of
the noncompact model are the $y_j$, the measure
in terms of the $Z_j$ variables is logarithmic.
To obtain the period of the quintic, we take a
derivative with respect to $t$. Thus, on the
corresponding cycles, we have
\begin{align}
 \Pi_{\rm nc}(t)&\propto
    \int \left(\bigwedge_j\frac{dZ_j}{Z_j}\right)
    e^{-w_{\rm HV}(Z)},
 \nonumber\\
 \Pi_{\rm c}(t)&\propto\partial_t\Pi_{\rm nc}(t)
    \propto e^{t/5}
    \int\left(\bigwedge_jdZ_j\right)e^{-w_{\rm HV}(Z)}.
 \label{eq:HVperiodmeasure}
\end{align}
Indeed, only the product term in the superpotential
depends on $t$. Taking the derivative brings down
a factor proportional to $e^{t/5}\prod_jZ_j$, which
cancels the factors of $Z_j^{-1}$ in the measure.
In this way, the logarithmic measure of the
noncompact model becomes the linear measure of
the compact model~\cite[Sections 6.3 and 7.1]{HV}.
We have included the overall sign from the derivative
in the normalization of the period.

As we discussed above, this difference can also be
understood from the R-charges of the original matter
fields. More generally, for a matter field of nonzero
R-charge $R_\Phi$, if we introduce the variable
\[
    Z=e^{-R_\Phi y/2},
\]
then
\[
    dZ=-\frac{R_\Phi}{2}e^{-R_\Phi y/2}dy.
\]
Thus, up to a constant, this variable has a linear
measure in the localized sphere
integral~\cite[Section 2 and Appendix B]{GS}.
Of course, to identify the mirror theory globally,
one also needs to specify the domain of the new
variable and its discrete identifications.

\subsubsection{The inverse-power potential in inverted coordinates}

Let us now consider the six-monomial MP
construction for the quintic, which will be
discussed in Section~\ref{subsec:quinticexample}.
On the hatted side, the $\widehat u_j$ fields
have charge one and $\widehat p$ has charge minus
five under the connected gauge group.

To compare the superpotential
$\widetilde w_-=\widetilde W_-|_0$ obtained from
the MP calculation with the Hori--Vafa mirror,
we introduce the variables
\begin{equation}
 Y_j=\widehat p^{-1/5}\widehat u_j^{-1}.
\end{equation}
Here we have chosen a fifth root on the dense
torus. Note that the two factors in $Y_j$ have
opposite gauge charges, so that $Y_j$ is invariant
under the connected gauge group.

Using the monomials in \eqref{eq:B}, we can write
the inverse-power superpotential in terms of
the $Y_j$ variables as
\begin{equation}
 \widetilde w_-=\sum_i\frac{a_i}{b_i}
       =\sum_{j=1}^5a_jY_j^5+a_0\prod_jY_j.
\end{equation}
Thus, at the level of the superpotential, we have
obtained precisely the polynomial we wanted.

However, as we learned from the Hori--Vafa
construction, to obtain the period of the compact
model, we need the linear measure in the $Y_j$
variables. The relation between the linear and
logarithmic measures is
\begin{equation}
 \bigwedge_jdY_j=\delta_0\,\Omega_{\log},\qquad
 \Omega_{\log}=\bigwedge_jd\log Y_j,\qquad
 \delta_0=\prod_jY_j=b_0^{-1}.
 \label{eq:delta}
\end{equation}
The question is therefore whether the integration
process in the original MP parent theory generates
this additional factor $\delta_0$ for the unit-state
period.

There is also a subtlety in the choice of the
integration cycle. The above field transformation
is well defined on the dense torus, but the
critical point $Y_1=\cdots=Y_5=0$ of the proposed
LG model is not a point in this torus. Hence,
even if we obtain the expected measure locally,
we still need to understand how the integration
cycle of the MP theory is related to that of the
compact mirror.

\subsubsection{The \texorpdfstring{$\mathcal P$}{P}-field determinant and the compact period measure}

Let us first see which factor is needed in the
integration measure to obtain the compact period
from the MP mirror. With the gauge choice
$\widehat p=1$, we have $Y_j=\widehat u_j^{-1}$,
and hence
\[
 dY_j=-\frac{1}{\widehat u_j^2}d\widehat u_j.
\]
Taking the wedge product over the five coordinates,
we obtain
\begin{equation}
 \bigwedge_{j=1}^5dY_j
 =-\delta_0^2\Omega_{\rm lin},
 \qquad
 \Omega_{\rm lin}=\bigwedge_{j=1}^5d\widehat u_j.
\end{equation}
Thus, relative to the linear reference measure
$\Omega_{\rm lin}$ in the original hatted variables,
we need a factor of $\delta_0^2$ to obtain the
linear measure in \eqref{eq:delta}.
Let us see whether this factor can arise from
integrating out the ordinary matter fields in
the MP parent theory.

Recall that the $\mathcal P$-field has R-charge two
in the compact model and zero in the noncompact
model. On the A-twisted two-sphere, keeping the
gauge flux and the Coulomb mass $m_0$ fixed,
changing its R-charge from zero to two decreases
the Dolbeault index by two. The fixed-flux one-loop formula
of~\cite[eq.~(2.38)]{CMP} then gives
\[
 \frac{Z_{\mathcal P}^{R=2}}{Z_{\mathcal P}^{R=0}}
 =m_0^2.
\]
At the formal Coulomb solution used in the MP
calculation for the mirror, we have
\[
 m_0=\frac{a_0}{b_0}=a_0\delta_0.
\]
The change of R-charge thus gives a factor
proportional to $\delta_0^2$ in the sphere
calculation. One might ask whether this is
precisely the factor needed in the period integral.

We also need to take into account the integration
over the six ordinary Coulomb scalars.
As explained in Appendix~\ref{app:measure},
the matter contribution to the curvature coupling
is proportional to
\[
 e^{2\pi i\Omega_\chi}
 \propto\frac{\prod_{i=0}^5m_i}{m_0^2},
\]
whereas the determinant of the Coulomb Hessian
is proportional to $(\prod_{i=0}^5m_i)^{-1}$.
Including the mixed derivatives with respect to
the hatted fields, we find
\[
 \mathcal H_{\rm ref}\propto
 \frac{1}{m_0^2}
 \det\left(
 \frac{\partial^2\widetilde w_-}
 {\partial\widehat u_j\partial\widehat u_k}
 \right).
\]

Recall that the handle-gluing operator of an LG
model with holomorphic volume form
$v(\widehat u)\Omega_{\rm lin}$ is given by
$\det w''/v(\widehat u)^2$ at a nondegenerate
critical point. Indeed, in coordinates with unit volume density the handle is
the Hessian determinant~\cite[eqs.~(2.9)--(2.10)]{GuLG}.
At a critical point the Hessian transforms as a bilinear form,
so its determinant acquires the square of the coordinate
Jacobian. The sphere contribution in the $\widehat u$
coordinates is therefore proportional to $v(\widehat u)^2/\det w''$.
Comparing the two expressions, we formally
associate the volume form
\begin{equation}
 m_0\,\Omega_{\rm lin}
\end{equation}
with the above calculation, up to a
field-independent normalization.
The factor of $m_0^2$ in the sphere calculation
corresponds to one power of $m_0$ in the volume
form because the latter enters quadratically
in the sphere correlator.

In terms of the $Y_j$ variables, we have
\[
 d\widehat u_j=-Y_j^{-1}d\log Y_j,
\]
and therefore
\begin{equation}
 \Omega_{\rm lin}=-\delta_0^{-1}\Omega_{\log},
 \qquad
 m_0\Omega_{\rm lin}=-a_0\Omega_{\log}.
 \label{eq:MPmeasuretest}
\end{equation}
Here $a_0$ is independent of the integration
variables, although it may depend on the parameters.
With our choice of reference measure, we obtain
the logarithmic measure of the noncompact mirror
rather than the linear measure of the compact
mirror.

Note that for generic coefficients, the homogeneous
quintic superpotential has no critical point on
the dense torus. Even if there is a critical point,
the homogeneity of the superpotential implies that
it is not isolated. The comparison with the
handle-gluing formula is therefore only a formal
calculation on the dense torus with $\widehat p=1$.

There is also the question of the reference measure
itself. We have chosen $\Omega_{\rm lin}$ rather
than deriving it from the MP path integral.
If the full calculation at $R=0$ gives the
logarithmic measure and the relative factor $m_0$
found above survives the remaining integrations,
we would obtain precisely the factor $\delta_0$
required to relate the noncompact period to the
compact one, up to a parameter-dependent
normalization.

We still need to take into account the mixed
gauge sector, the dynamical couplings
$U_i\log\Phi_i$, the remaining gauge sums,
and the integration cycle.
It is not clear to us whether the full MP
integral gives the measure required for the
compact period.
In the next section, we will instead modify the
matter content of the MP parent theory to obtain
positive powers of $B_i$ directly from the
one-loop calculation.

\section{The graded parent theory}
\label{sec:Higgs}
In this section, we propose a modification of the MP
parent theory. The idea is to reverse the sign of the
one-loop contribution by introducing Grassmann-odd
chiral superfields, while allowing the additional
fields to become massive in a Higgs vacuum. We will see how this modification leads to the
positive powers of the hatted monomials needed for
the Batyrev--Borisov superpotential.

\subsection{Reversing the one-loop contribution}

From the discussion in the previous sections, we know that
we cannot simply change the relative sign of the logarithmic
couplings in the MP parent theory, since this sign is fixed
by gauge invariance. However, one may ask whether we can
change the sign of the one-loop correction by modifying
the matter content of the theory.

Let us first consider one column of $P$. Recall that
integrating out the chiral superfield $\Phi$ of charge one
under $U(1)_H$ gives a contribution
\[
    -\log(\Sigma_H/\mu)
\]
to the derivative of the effective twisted superpotential
with respect to $\Sigma_H$~\cite[eq.~(3.39)]{HV}.
For a Grassmann-odd chiral superfield of the same charge,
the functional determinant is inverted, so the contribution
changes sign~\cite[Sec.~3.2]{GuZou}. Thus, if we introduce two
such fields $\Theta_1$ and $\Theta_2$, each of charge one
under $U(1)_H$, the contribution from these three fields
is
\[
    -\log(\Sigma_H/\mu)
    +2\log(\Sigma_H/\mu)
    =\log(\Sigma_H/\mu).
\]
In this way, we can change the sign of the one-loop
contribution without changing the logarithmic coupling
in the MP action.

Of course, introducing these additional matter fields
may change the IR physics of the original theory.
Since we would like to recover the original GLSM by
integrating out the additional fields, we need a vacuum
in which these fields are massive while the $U(1)_H$
gauge symmetry remains unbroken.

For this purpose, we introduce another Grassmann-odd
chiral superfield $\Xi_I$ for each $\Theta_I$, with
opposite charge under $U(1)_H$. We can then introduce
a superpotential coupling of the form
\[
    S\Theta_I\Xi_I,
\]
where $S$ is neutral under $U(1)_H$. If $S$ acquires
a nonzero vacuum expectation value, both $\Theta_I$
and $\Xi_I$ become massive without breaking the
$U(1)_H$ gauge symmetry.

We further introduce an auxiliary gauge group $U(1)_K$
under which $S$ has charge one. We take $\Theta_I$
to be neutral and assign charge minus one to $\Xi_I$
under $U(1)_K$, so that the above superpotential is
also gauge invariant under $U(1)_K$.

Note that the introduction of $\Xi_I$ also affects
the one-loop calculation on the mirror side.
Let $x=\Sigma_H|_0$ and $y=\Sigma_K|_0$. The Coulomb
mass of $\Xi_I$ is $-(x+y)$. Thus, when we perform
the reverse calculation, we need to take into account
both $\Sigma_H$ and $\Sigma_K$.

To complete the auxiliary sector, we introduce two
ordinary chiral superfields $T_I$ of charge minus one
under $U(1)_K$ and two neutral chiral superfields $N_I$.
By introducing a superpotential coupling $ST_IN_I$,
we can also make these fields massive in the vacuum
where $S$ has a nonzero expectation value.

The matter content and the charge assignments for
one column are summarized in the following table.

\begin{center}
\begin{tabular}{lcccc}
\toprule
Multiplet & Statistics of lowest component
          & $Q_H$ & $Q_K$ & trial $R$\\
\midrule
$\Phi$ (original) & even & $1$ & $0$ & $R_\Phi$\\
$\Theta_I$ & odd & $1$ & $0$ & $r_\Theta$\\
$\Xi_I$ & odd & $-1$ & $-1$ & $2-r_\Theta$\\
$S$ & even & $0$ & $1$ & $0$\\
$T_I$ & even & $0$ & $-1$ & $1$\\
$N_I$ & even & $0$ & $0$ & $1$\\
\bottomrule
\end{tabular}
\end{center}

Here $I=1,2$. By a Grassmann-odd chiral superfield,
we mean that its lowest component is Grassmann odd,
while its spinor components are Grassmann even.

The parameter $r_\Theta$ denotes the trial vector R-charge of
either $\Theta_I$. For the $i$-th column of $P$, we choose
\[
    R_\Phi=R_i=2\mu_i,\qquad r_{\Theta,i}=1+\mu_i,
\]
where $\mu_i$ was defined in \eqref{eq:BBgrading}.
Thus $R_i=0$ and $r_{\Theta,i}=1$ for a coordinate
column, while $R_i=2$ and $r_{\Theta,i}=2$ for a
$\mathcal P$-field column. The additional matter
fields in column $i$ are neutral under the hatted
gauge groups and under the gauge groups associated
with the other columns.

In addition to the kinetic terms for these fields
and the gauge kinetic and FI terms for $U(1)_{K_i}$,
we introduce the following ordinary superpotential:
\begin{equation}
 W_{\rm aux}=\sum_{i,I}
 \left(
 \lambda_iS_i\Theta_{iI}\Xi_{iI}
 +h_iS_iT_{iI}N_{iI}
 +\frac{M_i}{2}N_{iI}^2
 \right).
 \label{eq:Waux}
\end{equation}
Throughout our discussion, we take $\lambda_i$,
$h_i$ and $M_i$ to be nonzero. We denote the
holomorphic FI scale associated with $U(1)_{K_i}$ by
\[
    \kappa_i=\mu e^{2\pi i\tau_{K_i}(\mu)},
\]
where the constant phases arising in the one-loop
calculation are absorbed into $\kappa_i$.

Note that we have not introduced any new couplings
between the additional matter fields and the
monomials $B_i$ on the hatted side. All the dependence
on $B_i$ still comes from the original MP interaction.

In what follows, we refer to the calculation in which
we recover the original GLSM as the A-side calculation,
and to the reverse calculation as the B-side calculation.

\subsection{Action, couplings and R-charges}

Let us now write down the action of the extended parent
theory. For the $i$-th column, we denote the collection
of additional matter fields by
\[
    \mathcal F_i=
    \{\Theta_{iI},\Xi_{iI},S_i,T_{iI},N_{iI}\},
    \qquad I=1,2,
\]
where there is only one $S_i$ for each column.

The additional D-term and the twisted superpotential
associated with the FI parameters of the auxiliary
gauge groups are given by
\begin{align}
 \Delta K
 &=\sum_i\left(
 \sum_{Z\in\mathcal F_i}
 \overline Z
 e^{2Q_H(Z)V_i+2Q_K(Z)V_{K_i}}Z
 -\frac{|\Sigma_{K_i}|^2}{4e_{K_i}^2}
 \right),
 \nonumber\\
 \Delta\widetilde W_{\rm FI}
 &=\frac{i}{\sqrt2}
   \sum_i\tau_{K_i}\Sigma_{K_i}.
 \label{eq:auxD}
\end{align}

The full action is obtained by adding $\Delta K$
to the D-density of the original MP theory and
integrating over the full superspace. Recall that
the MP D-density already contains the kinetic terms
of the original matter and vector multiplets as well
as the mixed $V\widehat V$ term in \eqref{eq:MP}.
Thus, we do not introduce another kinetic term for
$\Phi_i$ in \eqref{eq:auxD}. We also include the
ordinary and twisted superpotential terms in
\eqref{eq:MP}, \eqref{eq:Waux} and \eqref{eq:auxD},
together with their complex conjugates.

Each auxiliary gauge group $U(1)_{K_i}$ has its own
gauge coupling and flux lattice. If we treat the
ordinary gauge multiplets as background fields,
the additional matter fields belonging to different
columns do not couple to each other. We can therefore
study the auxiliary system for each column separately.

However, this does not mean that the full parent
theory is a product of these systems. The columns
are still coupled through the original MP interaction.
For example, recall that the dynamical coefficients
$U_i$ of the logarithmic couplings satisfy
$\ell^TU=0$ for any $\ell\in\ker_{\mathbb Z}P$.
We need to take these couplings into account when
we study the remaining gauge integrations.

Let us also check the running of the FI parameters.
Recall that for a theory containing both ordinary
and Grassmann-odd chiral superfields, the one-loop
coefficient is~\cite[Sec.~3.1]{GuZou}
\[
    b_{\rm FI}
    =\sum_{\rm even}Q-\sum_{\rm odd}Q.
\]
For $U(1)_K$, the sum of the charges of the ordinary
matter fields is $1-2=-1$, while the sum of the
charges of the Grassmann-odd matter fields is $-2$.
Thus,
\[
    b_{{\rm FI},K}=(-1)-(-2)=1.
\]
For $U(1)_H$, the charges of the $\Theta_I$ and
$\Xi_I$ fields cancel each other, and we are left
with the contribution from the original field
$\Phi$. Hence,
\[
    b_{{\rm FI},H}=1.
\]
We see that both holomorphic FI scales carry one
power of the subtraction scale.

From the R-charge assignments in the table, one
can easily check that each term in the superpotential
\eqref{eq:Waux} has vector R-charge two. We will
use these R-charges when computing the indices of
the matter fields.

Note, however, that the nonvanishing signed sum of
the $U(1)_K$ charges also implies a mixed gauge--R
anomaly for the candidate axial R-current at fixed
FI scale. Thus, although we have specified the vector
R-charges of the microscopic fields, we have not yet
identified the superconformal R-current of the full
graded parent theory.

\subsection{Applying the extension to the full monomial matrix}

So far, we have mainly discussed the additional
fields associated with one column of $P$. For a
general Batyrev--Borisov construction, we introduce
the same auxiliary system for every column,
including the $r$ columns associated with the
$\mathcal P_\alpha$ fields.

Note that the gauge invariance of the superpotential
\eqref{eq:Waux} only depends on the charges under
$U(1)_{H_i}$ and $U(1)_{K_i}$. The construction
does not depend on the number of rows of $P$ or
the degrees of the equations defining the
Calabi--Yau manifold. Thus, we can introduce the
same auxiliary systems for a general rectangular
matrix $P$.

One may ask whether we need to change the sign of
the one-loop contribution in every column. In fact,
the R-charge condition alone does not necessarily
require us to do so.

Suppose that after the reverse calculation we
obtain a monomial $B_i^{\varepsilon_i}$ for each column,
where $\varepsilon_i=\pm1$. Let $\boldsymbol\nu_\varepsilon$ be the vector of
half R-charges of the hatted matter fields. Recall
that the weight of $B_i$ is $(P^T\boldsymbol\nu_\varepsilon)_i$.
If we require each $B_i^{\varepsilon_i}$ to have weight one,
we obtain
\begin{equation}
 P^T\boldsymbol\nu_\varepsilon=\varepsilon,\qquad
 \ell\cdot\varepsilon=0\quad
 \hbox{for every }\ell\in\ker_{\mathbb Z}P.
 \label{eq:signgrading}
\end{equation}
The second condition follows directly from the
first one, since
\[
    \ell^T\varepsilon=\ell^TP^T\boldsymbol\nu_\varepsilon
    =(P\ell)^T\boldsymbol\nu_\varepsilon=0.
\]

Recall from \eqref{eq:BBgrading} that the vector
$\boldsymbol\nu$ associated with the dual nef partition satisfies
$P^T\boldsymbol\nu=\mathbf 1_n$. Thus, choosing $\varepsilon_i=1$ for
all $i$ is consistent with the R-charge condition.

For some matrices $P$, there may also be other
choices of signs satisfying \eqref{eq:signgrading}.
However, in this paper we are interested in
recovering the particular mirror superpotential
determined by the dual nef partition. We therefore
choose the positive power of $B_i$ in every column.

In Section~\ref{subsec:quinticexample}, we will see
that the condition \eqref{eq:signgrading} gives a
stronger restriction on the possible sign choices
for the quintic.

\section{Two local reductions of the same parent}
\label{sec:reductions}

As in the Morrison--Plesser construction, we consider two
different ways of integrating out massive fields in our parent
theory. On the A side, the additional fields become massive in
the Higgs vacuum of the auxiliary gauge group, leaving the
local couplings of the MP theory. On the B side, we
keep the hatted matter fields fixed and integrate out the
ordinary matter fields. The relation between these two
calculations will be further studied in Section~\ref{sec:conjecture}.

\subsection{The \texorpdfstring{$S$}{S}-Higgs vacuum on the A side}
\label{sec:forward}

\subsubsection{The local mass spectrum}

Consider a positive FI parameter $\xi_K$ for $U(1)_K$.
We have a Higgs vacuum with $S|_0=v\ne0$. The F-term
equations for $T_I$ and $N_I$ give
$T_I|_0=N_I|_0=0$, and the D-term equation on the bosonic
body gives $|v|^2=\xi_K$. The phase of $v$ can be fixed
by a $U(1)_K$ gauge transformation.

In this vacuum, the two Grassmann-odd pairs have a
holomorphic mass $\lambda v$. For each pair of ordinary
chiral superfields $T_I$ and $N_I$, the mass matrix is
\begin{equation}
 \mathcal M_{TN}=
 \begin{pmatrix}
 0&hv\\
 hv&M
 \end{pmatrix},
 \qquad
 \det\mathcal M_{TN}=-(hv)^2.
 \label{eq:Higgsmass}
\end{equation}
The $U(1)_K$ vector multiplet also becomes massive, with
a mass of order $e_K|v|$. Since $S$ has charge one under
$U(1)_K$, there is no residual gauge symmetry associated
with $K$. On the other hand, the $U(1)_H$ gauge symmetry
is not broken by the expectation value of $S$.

With canonical kinetic terms at the vacuum, the physical
masses are the singular values of \eqref{eq:Higgsmass},
\begin{equation}
 m_\pm=
 \frac{\sqrt{|M|^2+4|hv|^2}\,\pm |M|}{2}.
 \label{eq:masses}
\end{equation}
Note that if we take $|M|\gg|hv|$, one of the fields
becomes light, with a mass of order $|hv|^2/|M|$.
Thus, instead of taking a large mass limit for $M$,
we keep $v,\lambda,h,M,e_K$ finite and nonzero and
consider the theory at an energy scale
\[
 E\ll\min(m_-,|\lambda v|,e_K|v|).
\]
Assuming a regular kinetic metric, we can integrate out
all the additional fields near this vacuum.

Since $\Theta_I$ and $\Xi_I$ have opposite charges under
$U(1)_H$, their contributions to the logarithmic
renormalization of the $H$-FI parameter cancel.
There may be a constant shift of the theta angle,
depending on the phases of the masses, which can be
absorbed into the definition of the holomorphic FI
parameters.

\subsubsection{Vortex cores and the original GLSM}

We should also consider the vortex contribution. The additional fields are massive in the Higgs vacuum, and some of their mass terms depend on $S$. However, $S$ can vanish at
isolated points on the worldsheet in a vortex background. So we need to understand what happens to
these mass terms at the vortex cores.

Consider a supersymmetric vortex in the A-twisted theory on a
compact Riemann surface $\mathcal C$. Let $L_K$ be the
$U(1)_K$ gauge line bundle of degree $k$. Since $S$ has charge
one under $U(1)_K$ and R-charge zero, its lowest component
$S_{\rm bos}=S|_0$ is a section of $L_K$. The BPS equation
$\bar\partial_{A_K}S_{\rm bos}=0$ makes this section holomorphic.
In the $S$-Higgs sector $S_{\rm bos}$ is not identically zero,
so its zeros define an effective divisor $D$ of degree $k$,
counted with multiplicity. In particular, $k\geq0$ in this sector.

For the odd fields, let $E_\Theta$ and $E_\Xi$ denote the bundles
of $\Theta_I$ and $\Xi_I$, respectively. When their R-charges are
odd, these bundles involve $K_{\mathcal C}^{1/2}$, so we choose a spin structure on $\mathcal C$. This choice is unique on the sphere. At higher genus, the Dolbeault indices do not depend on the choice of spin structure.

The gauge charges and R-charges in $S\Theta_I\Xi_I$ give
\[
  E_\Xi\otimes L_K
  \simeq K_{\mathcal C}\otimes E_\Theta^{-1}.
\]
Multiplication by $\lambda S_{\rm bos}$ then gives the short exact
sequence of sheaves
\begin{equation}
  0\longrightarrow E_\Xi
  \xrightarrow{\ \lambda S_{\rm bos}\ }
  K_{\mathcal C}\otimes E_\Theta^{-1}
  \longrightarrow
  (K_{\mathcal C}\otimes E_\Theta^{-1})|_D
  \longrightarrow 0.
  \label{eq:coresheaf}
\end{equation}
For $k=0$, $S_{\rm bos}$ is nowhere zero and the mass map
identifies the two bundles. For $k>0$, the map fails to be
invertible at the zeros of $S_{\rm bos}$. Its cokernel is
supported at the vortex cores and has length $k$ for each pair
$(\Theta_I,\Xi_I)$.

For the even fields, we can first eliminate $N_I$ from the
superpotential. Since $M\ne0$, its F-term equation gives
$N_I=-hST_I/M$. Substituting this back into the superpotential,
we find
\[
  W_{\rm aux,even}
  =-\frac{h^2}{2M}\sum_I S^2T_I^2.
\]
Thus the holomorphic mass of $T_I$ is proportional to
$S_{\rm bos}^2$ and has a double zero at each simple zero of
$S_{\rm bos}$. We can also see this from the mass matrix in
\eqref{eq:Higgsmass}: replacing $v$ with $S_{\rm bos}$ gives
$\det\mathcal M_{TN}=-h^2S_{\rm bos}^2$.

The additional fields may therefore contribute at the vortex cores, even though they are massive in the vacuum. We study their contribution to the vortex sum in Section~\ref{sec:vortex}, where we keep the ordinary gauge
multiplets fixed as backgrounds.

Let us now return to the calculation near the Higgs vacuum.
After integrating out the additional massive fields, we
recover the local holomorphic couplings of the MP parent
theory, up to the constant FI shifts discussed above. We
can then use Morrison and Plesser's argument~\cite[Sec.~5]{MP}
in what they call the
``confining phase.'' We first integrate out the massive
hatted matter and then solve the equations for the hatted
field-strength superfields. With the constraint $P\Sigma=0$
discussed in Section~\ref{subsec:MPgaugeReduction}, the surviving
gauge Lie algebra is $\ker_{\mathbb R}P$. The superpotential becomes
\begin{equation}
 W_A=\sum_{a=1}^{u}c_aA_a
     =\sum_{\alpha=1}^r
       \mathcal P_\alpha G_\alpha(\mathcal U).
 \label{eq:WAgeneral}
\end{equation}

In a geometric phase, the equations for the $\mathcal P_\alpha$
fields impose $G_\alpha=0$. The full F-term equations of
$w_A=W_A|_0$ are
\[
 G_\alpha(u)=0,\qquad
 \sum_{\alpha=1}^r p_\alpha\,dG_\alpha(u)=0.
\]
If the complete intersection is smooth, the $dG_\alpha$ are
independent on their common zero locus. The second equation
then requires all $p_\alpha$ to vanish. After taking the
gauge quotient, we obtain the compact Calabi--Yau manifold $Y$.

To obtain the full GLSM, we still have to recover the group
$G$ in \eqref{eq:BBgroups} from the parent theory. This includes
its finite components and the sum over gauge bundles.
The vortex contributions discussed above must also be taken into account. We choose the FI parameters as in
Section~\ref{subsec:BBintegral} and work out the quintic
example in Section~\ref{subsec:quinticexample}.

\subsection{The effective superpotential on the mirror side}
\label{sec:reverse}
We keep the microscopic couplings fixed and compute the effective superpotential on the regular massive branch.

\subsubsection{The signed determinant and Coulomb equations}

On the B side, we integrate out the ordinary matter
fields in the presence of the additional fields
introduced in Section~\ref{sec:Higgs}. Consider one
column of $P$ and set the hatted field strengths to
zero. We keep the hatted matter fields fixed and use
the following notation:
\[
 x=\Sigma_H|_0,\qquad
 y=\Sigma_K|_0,\qquad
 s=x+y,\qquad
 b=B|_0,\qquad
 q=\frac{a}{b}.
\]
The Coulomb masses of the charged matter fields can
be read from their gauge charges:
\begin{align*}
 m_\Phi=m_{\Theta_1}=m_{\Theta_2}&=x,\\
 m_{\Xi_1}=m_{\Xi_2}&=-(x+y),\\
 m_S&=y,\\
 m_{T_1}=m_{T_2}&=-y.
\end{align*}
The neutral fields $N_I$ have a superpotential mass $M$.
Around the Coulomb background with vanishing matter scalars,
the two cubic terms in \eqref{eq:Waux} start at third order
in fluctuations. They therefore do not enter the Gaussian
mass matrix.
For $xy(x+y)\ne0$, we can integrate out these fields
at an energy scale below their masses. In this
calculation, the hatted gauge multiplets are held
fixed, with $U_i=0$ on the chosen background.

To write down the one-loop effective twisted
superpotential, we introduce the notation
\[
 \mathcal L(\mathcal M)
 =\mathcal M[\log(\mathcal M/\mu)-1].
\]
Recall that integrating out an ordinary chiral
superfield gives $-\mathcal L(\mathcal M)$, while
a Grassmann-odd chiral superfield gives
$+\mathcal L(\mathcal M)$~\cite[Sec.~3.2]{GuZou}.

The contributions from $\Phi$ and the two $\Theta_I$
fields are
\[
 -\mathcal L(\Sigma_H)+2\mathcal L(\Sigma_H)
 =\mathcal L(\Sigma_H).
\]
The two $\Xi_I$ fields contribute
\[
 2\mathcal L(-\Sigma_H-\Sigma_K)
 =-2\mathcal L(\Sigma_H+\Sigma_K),
\]
up to terms linear in the field strengths. Similarly,
the contribution from $S$ and the two $T_I$ fields is
\[
 -\mathcal L(\Sigma_K)-2\mathcal L(-\Sigma_K)
 =\mathcal L(\Sigma_K),
\]
where we have again omitted the linear terms.

These linear terms depend on the choice of branches
of the logarithms and can be absorbed into the FI
parameters. Including the tree-level couplings, we
obtain the effective twisted superpotential
\begin{align}
 \widetilde W_{\rm eff}(\Sigma_H,\Sigma_K;B)
 &=\Sigma_H\log\frac{a}{\mu B}
   +\Sigma_K\log\frac{\kappa}{\mu}
   \nonumber\\
 &\quad+\mathcal L(\Sigma_H)
       +\mathcal L(\Sigma_K)
       -2\mathcal L(\Sigma_H+\Sigma_K),
 \label{eq:oneloop}
\end{align}
where the common normalization factor $C$ has been
suppressed. Note that the logarithmic coupling
between $\Sigma_H$ and $B$ is the same as in the
original MP parent theory.

We denote the lowest component of
\eqref{eq:oneloop} by
$\widetilde w_{\rm eff}(x,y;b)$.
Taking derivatives with respect to $x$ and $y$ gives
\begin{equation}
 \partial_x\widetilde w_{\rm eff}
 =\log\frac{qx}{(x+y)^2},\qquad
 \partial_y\widetilde w_{\rm eff}
 =\log\frac{\kappa y}{(x+y)^2}.
 \label{eq:gradHK}
\end{equation}
The theta angles of $U(1)_H$ and $U(1)_K$ are periodic,
so in our normalization the two derivatives vanish
modulo $2\pi i\mathbb Z$. Exponentiating gives
\[
 s^2=qx=\kappa y.
\]
It follows that
\[
 x=\frac{s^2}{q},\qquad y=\frac{s^2}{\kappa}.
\]
Substituting these expressions into $s=x+y$, we have
\[
 s=s^2\left(\frac{1}{q}+\frac{1}{\kappa}\right).
\]
Since $s\ne0$ in our calculation, we obtain
\begin{align}
 s^2&=qx=\kappa y,\nonumber\\
 x_*&=\frac{q\kappa^2}{(q+\kappa)^2},\qquad
 y_*=\frac{\kappa q^2}{(q+\kappa)^2},\qquad
 s_*=\frac{\kappa q}{q+\kappa}.
 \label{eq:saddle}
\end{align}
Note that the solution becomes singular at
$q=-\kappa$. The solution $s=0$ of the algebraic
equations cannot be studied using
\eqref{eq:oneloop}, since some of the matter
fields become massless there.

From \eqref{eq:gradHK}, we can also compute the Hessian,
\begin{equation}
 \widetilde w_{\rm eff}''
 =
 \begin{pmatrix}
 1/x&0\\
 0&1/y
 \end{pmatrix}
 -\frac{2}{x+y}
 \begin{pmatrix}
 1&1\\
 1&1
 \end{pmatrix}.
 \label{eq:HKHess}
\end{equation}

There is a simple way to see the change of sign in
the resulting interaction without first evaluating
the full effective superpotential. Let
\[
 f(b)=\widetilde w_{\rm eff}(x_*,y_*;b).
\]
On a stationary branch of the logarithms, we have
\[
 \frac{\partial f}{\partial\log q}=x_*,
 \qquad
 \frac{\partial f}{\partial\log\kappa}=y_*,
\]
since the derivatives of $f$ through $x_*$ and $y_*$
vanish by the vacuum equations. Using $q=a/b$, we obtain
\[
 -b\frac{\partial f}{\partial b}
 =x_*=\frac{\kappa^2}{a}b+O(b^2),
\]
where $a$ and $\kappa$ are kept fixed. Thus, near $b=0$,
the leading field-dependent term in $f(b)$ is
$-(\kappa^2/a)b$, rather than an inverse power of $b$.

Note that this change cannot be obtained by simply
reversing the sign of the $H$ charges in the original
MP theory. Under a change of the gauge generator,
both $Q_H$ and $\Sigma_H$ change sign, and hence
the Coulomb mass $Q_H\Sigma_H$ remains unchanged.
One also needs to transform the FI coupling and the
gauge bundles accordingly. In our calculation,
the change comes from the one-loop contributions
of the additional matter fields.

For the above calculation to describe an integration
over massive fields, we assume a regular kinetic
metric for the Coulomb fields and an appropriate
integration prescription for the Grassmann-odd
multiplets.

\subsubsection{The rational threshold}

The determinant of the Hessian in \eqref{eq:HKHess}
is
\[
 \det\widetilde w_{\rm eff}''=-\frac{1}{xy}.
\]
Thus, the solution \eqref{eq:saddle} is isolated
in the region where the charged matter fields
are massive.

Substituting \eqref{eq:saddle} into
\eqref{eq:oneloop}, the logarithmic terms cancel
on the stationary branch, and we obtain
\[
 f(b)=x_*+y_*
     =\frac{\kappa q}{q+\kappa}
     =\frac{\kappa a}{a+\kappa b}.
\]
In terms of the monomial superfield $B$, the resulting
interaction can be written as
\begin{equation}
 F(B)=\frac{\kappa a}{a+\kappa B},\qquad
 F(B)-\kappa=\frac{dB}{1-dB/\kappa},\qquad
 d=-\frac{\kappa^2}{a}.
 \label{eq:threshold}
\end{equation}

Recall that in the original MP calculation,
the resulting superpotential is proportional to
$1/B$. In our case, after subtracting the
field-independent constant $\kappa$, we have
\[
 F(B)-\kappa=dB+O(B^2).
\]
We therefore obtain the desired positive power of
$B$, together with higher powers.

There are two issues that we still need to understand.
First, we need to determine whether these higher-order terms affect the IR
physics. Second, although $F(B)$ is regular at $B=0$,
the calculation leading to \eqref{eq:threshold} was
performed for $b\ne0$ and $a+\kappa b\ne0$.
Some of the matter fields that we have integrated out
become light as we approach $b=0$.

Furthermore, in the above calculation we have treated
the hatted gauge multiplets as backgrounds. To identify
the full mirror theory, we still need to take into
account the remaining gauge integrations in the
presence of the mixed MP interaction. We will return
to these issues in the following sections.

\section{The Batyrev--Borisov critical sector at finite couplings}
\label{sec:fixed}

\subsection{The general interaction and algebraic parameter matching}

Recall that the coefficient of the linear term in $B_i$
in \eqref{eq:threshold} is $d_i=-\kappa_i^2/a_i$.
Let us choose
\begin{equation}
 \kappa_i=Ld_i,\qquad a_i=-L^2d_i,\qquad
 0<L<\infty,\qquad d_i\ne0.
 \label{eq:ray}
\end{equation}
Here $L$ is a dimensionless parameter which we keep fixed
when taking the IR limit, and $\mu$ still denotes the
subtraction scale. Summing over all the columns of $P$
and subtracting the field-independent constants $\kappa_i$,
we obtain
\begin{align}
 W_L&=\sum_{i=1}^{n}\frac{d_iB_i}{1-B_i/L},
 &
 W_{\rm BB}&=\sum_{i=1}^{n}d_iB_i
      =\sum_{\beta=1}^r\widehat{\mathcal P}_\beta
       \widehat G_\beta(\widehat{\mathcal U}).
 \label{eq:BBWL}
\end{align}
The difference between the two superpotentials is
\[
 W_L-W_{\rm BB}=\sum_i\frac{d_iB_i^2}{L-B_i}.
\]
Since each $B_i$ is invariant under $\widehat G$,
this difference is also gauge invariant. Each $B_i$
is linear in one $\widehat{\mathcal P}_{\beta(i)}$,
so the difference is at least quadratic in the
$\widehat{\mathcal P}$-fields near $\widehat Y$.

For $|b_i|<L$ for all $i$, we can expand the difference as
\begin{equation}
 W_L-W_{\rm BB}=\sum_{k\ge2}L^{1-k}\sum_i d_iB_i^k.
 \label{eq:BBhigher}
\end{equation}
Of course, the rational superpotential itself is not
restricted to this region. In particular, $|b_i|=L$
does not necessarily give a singularity; the individual
terms can have poles at $b_i=L$. We will first study
the theory near $b_i=0$. The other regions of field
space need to be considered separately.

Let us now compare the parameters on the two sides.
From $d_i=-\kappa_i^2/a_i$ and the definition of the
coordinates in \eqref{eq:BBcharacters}, we have
\begin{equation}
 z_\ell(d)=\frac{z_\ell(\kappa)^2}{z_\ell(a)},\qquad
 \kappa_i=Ld_i,\quad a_i=-L^2d_i
 \quad\Longrightarrow\quad z_\ell(a)=z_\ell(d).
 \label{eq:BBmodulimatching}
\end{equation}
Here we have used the Calabi--Yau condition
$\sum_i\ell_i=0$ for $\ell\in\ker_{\mathbb Z}P$.
Thus, the factor $-L^2$ in $a_i=-L^2d_i$ does not
appear in $z_\ell(a)$. We obtain the expected relation
between the FI parameters of the original GLSM and
the coefficients of the mirror superpotential for any
finite $L$. Similarly, we choose the hatted FI parameters
according to $\widehat z_{\widehat\ell}(c)$.

So far, we have only compared the algebraic parameters
of the two GLSMs. It would also be interesting to
understand the relation between the corresponding
marginal operators and their correlation functions,
as well as the flat coordinates on the moduli spaces
of the two theories.

\subsection{The compact critical component in a geometric phase}
\label{subsec:BBgeometriccritical}

We would like to see whether the difference between
$W_L$ and $W_{\rm BB}$ affects the IR physics.
Consider a geometric phase of the mirror GLSM in which
\[
 \widehat Y=\{\widehat G_1=\cdots=\widehat G_r=0\}
\]
is a smooth compact complete intersection.

Locally on the base, we trivialize the inverse nef line bundles and use $\widehat p_\beta$ as fibre coordinates. Near an orbifold point, we do this on the local cover and keep track of the action of the finite group.

Grouping the $B_i$ by their common
$\widehat{\mathcal P}_{\beta(i)}$, we can rewrite $w_L=W_L|_0$ as
\begin{align}
 w_L&=\sum_{\beta=1}^r\widehat p_\beta
       \widehat G_{\beta,L}(\widehat u,\widehat p_\beta),
 \nonumber\\
 \widehat G_{\beta,L}
   &=\sum_{i:\,\beta(i)=\beta}
        \frac{d_i\widehat M_i(\widehat u)}
             {1-\widehat p_\beta\widehat M_i(\widehat u)/L}.
 \label{eq:BBdeformedsection}
\end{align}
Here $\widehat M_i(\widehat u)$ is the lowest component
of $\widehat M_i(\widehat{\mathcal U})$. Note that
\eqref{eq:BBdeformedsection} is an exact expression,
rather than an expansion in powers of $1/L$.
In particular,
\[
 \widehat G_{\beta,L}(\widehat u,0)
 =\widehat G_\beta(\widehat u).
\]
If we denote the ideal generated by the fibre coordinates
by $\mathcal I=(\widehat p_1,\ldots,\widehat p_r)$,
we have
\[
 w_L-w_{\rm BB}\in\mathcal I^2
\]
near $\widehat p=0$, where $w_{\rm BB}=W_{\rm BB}|_0$.

Since $\widehat Y$ is a smooth complete intersection,
the differentials $d\widehat G_\beta$ are independent
along $\widehat Y$. We can thus use $\widehat G_\beta$
as local coordinates in the normal directions.
Denote the remaining coordinates, which parametrize
$\widehat Y$, by $\widehat u_\parallel$. We extend them
to a neighborhood of $\widehat Y$.

At $\widehat p=0$, the derivatives of
$\widehat G_{\beta,L}$ with respect to $\widehat u$
are the same as those of $\widehat G_\beta$.
Hence, by the holomorphic inverse function theorem,
we can make the following change of coordinates:
\begin{equation}
 (\widehat u,\widehat p)\longmapsto
 \bigl(\widehat u_\parallel,\ z_\beta=\widehat G_{\beta,L}
          (\widehat u,\widehat p_\beta),\ \widehat p\bigr).
 \label{eq:BBlocalcoordinates}
\end{equation}
In these coordinates, the superpotential is simply
bilinear in $\widehat p_\beta$ and $z_\beta$.
In a sufficiently small neighborhood $U$ of a point
of $\widehat Y$, we have
\begin{equation}
 w_L=\sum_{\beta=1}^r\widehat p_\beta z_\beta,\qquad
 \operatorname{Crit}(w_L)\cap U
   =\{\widehat p=z=0\}\cap U.
 \label{eq:BBcriticalscheme}
\end{equation}
Thus, locally near $\widehat Y$, the difference
between $w_L$ and $w_{\rm BB}$ can be absorbed into
the definition of the normal coordinates.

We can also check the masses of these fields directly
in the original coordinates $(\widehat p,\widehat G)$.
The Hessian in the normal directions is
\begin{equation}
 \left.\operatorname{Hess}_{\perp}w_L\right|_{\widehat Y}
 =\begin{pmatrix}\mathsf H&I_r\\ I_r&0\end{pmatrix},
 \qquad
 \det\operatorname{Hess}_{\perp}w_L=(-1)^r.
 \label{eq:BBnormalHessian}
\end{equation}
The matrix $\mathsf H$ comes from the terms quadratic
in $\widehat p$. Exchanging the two blocks of $r$ columns
gives a triangular matrix with unit diagonal and explains
the sign $(-1)^{r^2}=(-1)^r$. Thus, if the kinetic metric
is regular, the $\widehat{\mathcal P}_\beta$ fields pair with
the normal fields and become massive. After integrating out these
fields in the reduced hatted theory, we locally recover
the target $\widehat Y$ with the expected complex
structure.

There is no need to take $L$ to infinity in the above
calculation. Of course, the kinetic terms of the
remaining fields can depend on $L$. Also, away from
the neighborhood of $\widehat Y$, the rational
superpotential may have other critical points.
We will see this explicitly for the quintic in
Appendix~\ref{app:extra}.

The same argument works for complete intersections
of any dimension and codimension, as long as the
defining equations meet transversely. In higher
dimensions, however, a crepant partial resolution
does not necessarily remove all the singularities
of the Calabi--Yau manifold~\cite{BBcomplete}.
Near such singularities, we need to study the
corresponding GLSM rather than use the above
change of coordinates.

\subsection{Landau--Ginzburg and hybrid descriptions}

If the GLSM has an LG or hybrid phase, we can also
study the difference between $W_L$ and $W_{\rm BB}$
from the R-symmetry of the IR theory.

Suppose that the critical sector under consideration
is compact and that each $B_i$ has R-charge two in the
mirror LG convention. If $B_i^k$ represents a nonzero
chiral primary, its dimension is half its R-charge,
$\Delta=k$~\cite{LVW}. The two supercharges in the
F-term descendant raise its dimension to $k+1$. Its coupling
has RG eigenvalue $1-k$. Thus, the terms with
$k\geq2$ in \eqref{eq:BBhigher} are irrelevant
deformations of this fixed point. This does not
exclude the existence of other critical points of
$W_L$ at finite $L$.

For an LG model with an isolated critical point,
one can explicitly check which of these higher-order
terms are nontrivial in the Jacobi ring. In a hybrid
model, one should also take into account the
contributions from the compact base and the twisted
sectors~\cite{BMPHybrid}. In applying the above argument
to a hybrid model, we have assumed the corresponding
IR R-symmetry.

Note that we did not need this assumption in our
discussion of the geometric phase. We will consider
the quintic LG model and the hybrid model for two
cubics in $\mathbb P^5$ in Section~\ref{sec:examples}.

\subsection{Light fields in the parent theory}

Although the rational superpotential is regular at $B_i=0$,
we should check the masses of the fields that were integrated
out. On the geometric branch of $\widehat Y$, all the
$\widehat{\mathcal P}_\alpha$ fields vanish. Since
$B_i=\widehat{\mathcal P}_{\beta(i)}
\widehat M_i(\widehat{\mathcal U})$, we have $b_i=0$
on $\widehat Y$. From \eqref{eq:saddle}, we find
\begin{equation}
  x_{i,*}=\frac{\kappa_i^2}{a_i}b_i+O(b_i^2),\qquad
  y_{i,*},\,s_{i,*}\longrightarrow\kappa_i
  \quad (b_i\to0).
  \label{eq:BBparentlight}
\end{equation}
Here $a_i$ and $\kappa_i$ are kept nonzero. Thus, the fields
with Coulomb masses $y_i$ and $s_i$ remain massive, but
$\Phi_i$, $\Theta_{i1}$ and $\Theta_{i2}$, whose Coulomb
masses are $x_i$, become light.

Holding $x_i$ fixed, take the solution of the $K_i$ equation
with $y_i\to\kappa_i$ as $x_i\to0$. The $K_i$-charged matter
is massive there, and $\partial_{y_i}^2\widetilde w_{\rm eff}
\to-1/\kappa_i$ by \eqref{eq:HKHess}. We can therefore eliminate
$y_i$ locally. To read off the holomorphic scale of the remaining
$H_i$ theory, take $x_i$ small and nonzero and include the
one-loop logarithm of its charged matter. Equation~\eqref{eq:gradHK} gives
$\partial_{x_i}\widetilde w_{\rm red}
=\log(x_i/\Lambda_{H_i})+O(x_i/\kappa_i)$, with
\[
  \Lambda_{H_i}(b_i)
  =\frac{\kappa_i^2}{a_i}b_i=-d_i b_i.
\]
This scale vanishes at $b_i=0$. In Appendix~\ref{app:local},
we also calculate the Jacobian of the Coulomb equations,
after clearing their denominators. It is nonzero at
$(b_i,x_i,y_i)=(0,0,\kappa_i)$, so the solution itself
can be continued to $b_i=0$. However, this does not mean
that we can still integrate out $\Phi_i$, $\Theta_{i1}$
and $\Theta_{i2}$ at this point.

A similar issue already appears in the original MP calculation.
On the A side, the stationary equation of MP~\cite[eq.~(12)]{MP}
gives $\widehat\sigma_a=c_a A_a|_0$ in our normalization.
Each monomial $A_a$ contains a $\mathcal P_\alpha$ field, so
$\widehat\sigma_a$ tends to zero
as we approach the Calabi--Yau locus, where all the
$\mathcal P_\alpha$ fields vanish. The hatted matter that MP
integrated out then becomes light. In our calculation on the
mirror side, the same issue arises for $\Phi_i$ and the two
Grassmann-odd fields $\Theta_{i1}$ and $\Theta_{i2}$.

Let us also consider the solution away from $b_i=0$. With the
parameter choice \eqref{eq:ray}, we can write \eqref{eq:saddle} as
\[
  x_{i,*}=-\frac{d_i b_i}{(1-b_i/L)^2},\qquad
  y_{i,*}=\frac{\kappa_i}{(1-b_i/L)^2}.
\]
Both Coulomb scalars go to infinity as $b_i\to L$. It is not
clear what this pole corresponds to in the full parent theory.
In the limit $|b_i|\to\infty$, both scalars tend to zero, so
the charged matter again becomes light. The expansion
\eqref{eq:BBhigher} around $b_i=0$ does not determine whether
these other regions of field space contribute to the parent
theory's path integral.

The mass matrix \eqref{eq:BBnormalHessian} describes the fields
of the reduced hatted theory near $\widehat Y$. To determine
whether this theory and the $S$-Higgs sector give mirror
descriptions of the same IR SCFT, we still need to understand
the dynamics of the light fields in the parent theory. In the
next section, we compare observables in the two descriptions
to test this proposed relation.

\section{Evidence for a common infrared sector}
\label{sec:conjecture}

In the previous sections, we have studied two ways of
integrating out massive fields in the extended parent theory.
In this section, we study the relation between the two
calculations by looking at the vortex contributions and the
operators associated with the FI parameters.

\subsection{Vortices in the \texorpdfstring{$S$}{S}-Higgs vacuum}
\label{sec:vortex}

\subsubsection{The vortex sum}

Let us fix a supersymmetric background for the $H$ gauge
multiplet and set the flavor masses and fluxes to zero.
The original MP matter fields are neutral under $U(1)_K$,
and the mixed D-term has no dependence on the $K$ gauge
field. We compute the contribution of the $S$ pole to the
$K$ vortex sum in this background.

We work with the equivariant A-twist, with rotation parameter
$\epsilon>0$. For the south-pole vortex contribution we hold
the north-pole vortex number at zero. Since $S$ has charge
one and R-charge zero, the $S$ pole then has $y_{\rm N}=0$
and $y_{\rm S}=n\epsilon$ in $K$ flux $n\geq0$
\cite[eqs.~(9.18)--(9.21)]{CCP}. We denote the $H$ flux
by $m_H$ and the average of the $H$ Coulomb scalars at
the two poles by $x$. Then at the south pole we have
\[
    x_{\rm S}=x+\frac{m_H\epsilon}{2}.
\]

We normalize the vortex sum by its contribution at $n=0$
and choose the phase of the FI parameter so that
$z_{\rm v}=\kappa/\epsilon$. The equivariant chiral
determinant~\cite[eqs.~(4.41)--(4.42)]{CCP} gives a factor
of $1/n!$ from the $S$ residue and $(1/2)_n$ from each $T_I$,
with the powers of $\epsilon$ included in $z_{\rm v}$.
Reversing the statistics in the same matter complex inverts
the determinant, as in the A-twisted sphere calculation
of~\cite[eqs.~(15a)--(15b)]{GuZou}. Thus each $\Xi_I$
contributes $1/(x_{\rm S}/\epsilon+R(\Xi_I)/2)_n$.
The determinants of the fields neutral under $U(1)_K$
cancel against their values at $n=0$. We obtain
\begin{equation}
 \mathcal V_S(x)=\sum_{n=0}^{\infty}w_n,\qquad
 w_n=\frac{z_{\rm v}^{\,n}}{n!}
       \left[\frac{(1/2)_n}{(x/\epsilon+\gamma)_n}\right]^2,
 \qquad
 \gamma=\frac12+\frac{m_H}{2}-\frac{R_\Phi}{4},
 \qquad w_0=1.
 \label{eq:vortex}
\end{equation}
Here $(\alpha)_n=\Gamma(\alpha+n)/\Gamma(\alpha)$, and
$R_\Phi=2\mu_i$ for the $i$-th column of $P$.
The R-charge is zero for the ordinary coordinate fields
and two for the $\mathcal P$-fields. We take $x$ away
from the poles of the determinants.

\subsubsection{The semiclassical limit}

Recall that the $K$ Coulomb equation in
\eqref{eq:saddle} is
\[
    (x+y)^2=\kappa y.
\]
For a given $x$, this equation can have two positive real
solutions. The question is which of them we should
associate with the $S$-Higgs vacuum.

For simplicity, let us consider positive real $x$ and $\kappa$,
and take $\epsilon\to0^+$ at fixed $x$, $\kappa$ and $m_H$.
In this case all the terms in the vortex sum are positive
when $\epsilon$ is sufficiently small. We define $u=\epsilon n$,
which means that $n$ grows as $\epsilon\to0$ with $u$ fixed.
From \eqref{eq:vortex} one can find
\begin{equation}
 \frac{w_{n+1}}{w_n}
 =\frac{\kappa(n+1/2)^2}
        {\epsilon(n+1)(n+x/\epsilon+\gamma)^2}
 \ \longrightarrow\ \frac{\kappa u}{(x+u)^2}.
\end{equation}
Because $w_0=1$, $w_n$ is the product of the ratios
from degree zero to degree $n-1$. Taking the logarithm
turns this product into a Riemann sum. For $\epsilon n\to u$
we obtain
\begin{align}
 \lim_{\epsilon\to0^+}\epsilon\log w_n
 &=\mathcal A_x(u)
   =\int_0^u\log\frac{\kappa v}{(x+v)^2}\,dv
 \nonumber\\
 &=u\log\frac{\kappa u}{(x+u)^2}
   +u-2x\log(1+u/x),
 \qquad \mathcal A_x(0)=0.
 \label{eq:rate}
\end{align}
The integral is well defined at its lower endpoint.
For sufficiently large $u$, the ratio of consecutive
terms is bounded above by a constant smaller than one.

The stationary equation $\partial_u\mathcal A_x(u)=0$ takes
the same form as the $U(1)_K$ Coulomb equation if we set
$u=y$. Here $u=\epsilon n$ comes from the vortex number,
while $y$ is a Coulomb scalar. We compare them using the
$K$-FI derivative. The vortex sum gives
$\epsilon\partial_{\log\kappa}\log\mathcal V_S
=\epsilon\langle n\rangle$, while the Coulomb FI coupling
$y\log(\kappa/\mu)$ gives $y$. For $\kappa>4x$, the two
roots are
\begin{equation}
 y_\pm(x)=
 \frac{\kappa-2x\pm\sqrt{\kappa(\kappa-4x)}}{2}.
 \label{eq:roots}
\end{equation}
Since
\[
    \mathcal A_x''(u)=\frac{x-u}{u(x+u)},
\]
one can see that $y_-$ is a local minimum while $y_+$ is a
local maximum. However, this local maximum does not
necessarily dominate the vortex sum, because we also have to
compare it with the term at $n=0$.

Their difference at the level of the exponential rate is
\begin{equation}
 \mathcal A_x(y_+)-\mathcal A_x(0)
 =x\big[v-2\log(1+v)\big],\qquad v=y_+/x.
 \label{eq:height}
\end{equation}
The equation $v=2\log(1+v)$ has a nonzero positive
solution $v_c=2.5128624\ldots$. Thus, for $v>v_c$,
the terms around $y_+$ dominate the vortex sum.
For $v<v_c$, the sum is dominated by the endpoint $u=0$
at the exponential level. At $v=v_c$, the two
contributions have the same exponential rate.

\begin{figure}[tbp]
 \centering
 \begin{tikzpicture}
 \begin{axis}[width=.82\linewidth,height=5.7cm,
  xlabel={vortex variable \(u/x\)},
  ylabel={rate \(\mathcal A_x(u)/x\)},
  xmin=0,xmax=5.4,ymin=-.94,ymax=.83,axis lines=left,
  xtick={0,1,2,3,4,5},ytick={-.5,0,.5},
  legend style={at={(.03,.98)},anchor=north west,
                draw=none,fill=white,font=\small},
  ticklabel style={font=\small},clip=true]
  \addplot[gray,densely dotted,forget plot]
       coordinates {(0,0)(5.4,0)};
  \addplot[linkblue,thick,domain=.0001:5.4,samples=201]
       {x*ln(6*x/((1+x)^2))+x-2*ln(1+x)};
  \addlegendentry{\(\kappa/x=6\)}
  \addplot[black,thick,dashed,domain=.0001:5.4,samples=201]
       {x*ln(4.4*x/((1+x)^2))+x-2*ln(1+x)};
  \addlegendentry{\(\kappa/x=4.4\)}
  \addplot[only marks,mark=*,mark size=2pt,
       color=linkblue,forget plot]
       coordinates {(3.73205081,0.62333344)};
  \addplot[only marks,mark=square*,mark size=1.8pt,
       color=black,forget plot]
       coordinates {(1.86332496,-0.24064209)};
 \end{axis}
 \end{tikzpicture}
 \caption{The function $\mathcal A_x(u)/x$ for
 $\kappa/x=4.4$ and $\kappa/x=6$.
 The points marked in the figure correspond to $y_+$.
 Although there are two positive Coulomb solutions in
 both cases, the vortex sum is dominated by $y_+$ only in
 the latter case.}
 \label{fig:rate}
\end{figure}
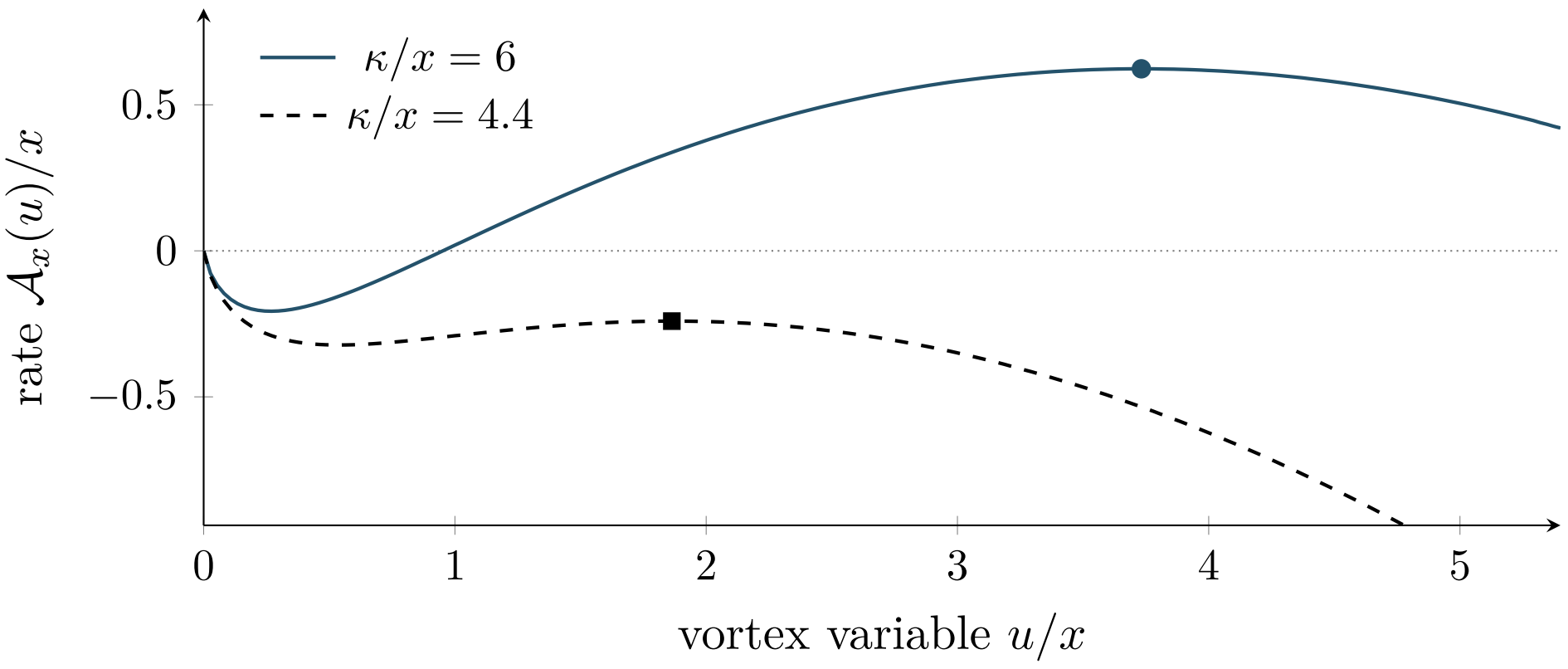

Note that two distinct Coulomb solutions exist when
$\kappa/x>4$, whereas the larger solution dominates the
vortex sum when
\[
    \frac{\kappa}{x}>v_c+2+\frac{1}{v_c}.
\]
We plot $\mathcal A_x(u)/x$ for two choices of
$\kappa/x$ in Figure~\ref{fig:rate}.

\subsubsection{An integral representation of the vortex sum}

One can also study the semiclassical limit by writing
the vortex sum as an integral. This lets us compare the
interior saddles with the boundary contribution. Let
$\mathfrak b=x/\epsilon+\gamma>1/2$ and define
\[
 d\nu_{\mathfrak b}(\xi)=
 \frac{\xi^{-1/2}(1-\xi)^{\mathfrak b-3/2}}
      {\mathrm B(1/2,\mathfrak b-1/2)}\,d\xi,
 \qquad \xi\in[0,1].
\]
We have
\[
    \int_0^1\xi^n\,d\nu_{\mathfrak b}(\xi)
    =\frac{(1/2)_n}{(\mathfrak b)_n}.
\]
Using this identity twice in \eqref{eq:vortex}, we obtain
\begin{equation}
 \mathcal V_S(x)=
 \int_0^1d\nu_{\mathfrak b}(\xi_1)
 \int_0^1d\nu_{\mathfrak b}(\xi_2)
 \exp(z_{\rm v}\xi_1\xi_2).
 \label{eq:vortexbeta}
\end{equation}
In the limit $\epsilon\to0^+$, the function appearing
in the exponent, including the contribution from the
measure and suppressing the factor $1/\epsilon$, is
\[
    \kappa \xi_1\xi_2+x\log(1-\xi_1)+x\log(1-\xi_2).
\]
The stationary equations in the interior give
\[
    \xi_1=\xi_2=\xi,\qquad x=\kappa \xi(1-\xi).
\]
If we set $y=\kappa \xi^2$, we recover the two solutions
in \eqref{eq:roots}, and the value of the function at
either solution is $\mathcal A_x(y)$. On the other
hand, along the edges $\xi_1=0$ or $\xi_2=0$, the maximum
is zero at $\xi_1=\xi_2=0$. Thus the two types of contributions
in the vortex sum are also present in this
integral.

The recurrence relation for $w_n$ also leads to a differential
equation for $\mathcal V_S$,
\begin{equation}
 D_{\rm v}[D_{\rm v}+x+(\gamma-1)\epsilon]^2\mathcal V_S
 =\kappa(D_{\rm v}+\epsilon/2)^2\mathcal V_S,
 \label{eq:vortexeq}
\end{equation}
where $D_{\rm v}=\epsilon z_{\rm v}\partial_{z_{\rm v}}$ and
$\kappa=\epsilon z_{\rm v}$. We substitute
$\mathcal V_S\sim\exp(\mathcal F_{\rm v}/\epsilon)$ and keep the
leading terms in $\epsilon$, which gives
\[
    y(x+y)^2=\kappa y^2,\qquad
    y=\partial_{\log\kappa}\mathcal F_{\rm v}.
\]
Dividing this equation by $y$ we get the Coulomb equation,
but the solution $y=0$ is lost. From the calculation above,
this solution can dominate the vortex sum for some
values of the parameters. Thus the differential equation
alone does not tell us which solution is selected.

\subsubsection{Comparison with the B-side calculation}

Let us consider the region $v>v_c$. As $w_n$ is proportional
to $\kappa^n$, we have $\partial_{\log\kappa}w_n=nw_n$
at fixed $x$ and $\epsilon$. In the semiclassical limit,
the dominant vortex number satisfies $\epsilon n\to y_+$,
and hence
\[
 \epsilon\,\partial_{\log\kappa}\log\mathcal V_S
 =\epsilon\langle n\rangle\longrightarrow y_+.
\]
In this region, $\mathcal F_{\rm v}(x,\kappa)=\mathcal A_x(y_+)$.
Taking a derivative with respect to $x$ gives
\[
    \partial_x\mathcal F_{\rm v}=-2\log(1+y_+/x),
\]
where we have used the stationary equation for $y_+$.
Thus, relative to its degree-zero normalization, the contribution
from the auxiliary sector to the exponentiated $H$ Coulomb equation is
\[
    e^{\partial_x\mathcal F_{\rm v}}=\frac{x^2}{(x+y_+)^2}.
\]
The factor $q/x$ comes from the $y$-independent part of
the B-side twisted superpotential. To see this, write
\[
 \widetilde w_K(x,y)=y\log(\kappa/\mu)
   +\mathcal L(y)-2\mathcal L(x+y),
 \qquad \mathcal L(z)=z[\log(z/\mu)-1].
\]
Since the vortex sum is normalized at $n=0$, we have
$\mathcal A_x(y)=\widetilde w_K(x,y)-\widetilde w_K(x,0)$.
The full twisted superpotential can then be written as
\[
 \widetilde w_{\rm eff}(x,y)
 =\underbrace{x\log(q/\mu)-\mathcal L(x)}_{\widetilde w_{H,0}(x)}
 +\mathcal A_x(y).
\]
The first term gives
$e^{\partial_x\widetilde w_{H,0}}=q/x$,
while $e^{\partial_x\mathcal F_{\rm v}}$ gives the vortex contribution.
The $H$ Coulomb equation therefore becomes
\[
    \frac{q}{x}\frac{x^2}{(x+y_+)^2}=1.
\]
This gives the $H$ Coulomb equation $qx=(x+y_+)^2$
used in \eqref{eq:saddle}.

From \eqref{eq:saddle}, we also have
$y_*/x_*=q/\kappa$. For real $q,\kappa>0$, the two roots
of the $K$ equation at $x=x_*$
satisfy $y_+y_-=x_*^2$. Thus $q/\kappa>v_c>1$ first
identifies $y_*$ as the larger root and then places it
above the endpoint in \eqref{eq:height}. We obtain
\begin{equation}
 q,\kappa>0,\quad q/\kappa>v_c
 \quad\Longrightarrow\quad
 \left.\lim_{\epsilon\to0^+}
 \epsilon\,\partial_{\log\kappa}
 \log\mathcal V_S(x)\right|_{x=x_*}
 =y_+(x_*)=y_*.
 \label{eq:bridge}
\end{equation}
Here the derivative is taken at fixed $x$.

For a general BB construction, one can choose the hatted
background fields such that the above condition holds
for all the columns of $P$. As an example, let us consider
$L>0$ and $d_i>0$. Since $P^T\boldsymbol\nu=\mathbf1_n$, we can
take
\begin{equation}
 \widehat\phi_a
 =\exp\!\bigl[\nu_a(\log s_0+i\pi)\bigr],
 \qquad b_i=-s_0,\qquad 0<s_0<L/v_c,
 \label{eq:generalvortexbackground}
\end{equation}
on a logarithmic cover. With the choice of parameters
in \eqref{eq:ray}, we have $q_i/\kappa_i=L/s_0>v_c$
for every $i$. At this background all the $x_{i,*}$
are nonzero, and we can take the energy to be below the
masses of the charged matter fields. In the vortex
calculation, the parameter that we take to zero is
$\epsilon$.

The vortex calculation thus supports the choice of the
B-side Coulomb solution in this region of parameters. In the
above calculation we have kept the $H$ gauge multiplets as
background fields. In the full parent theory, however, we also
integrate over these gauge multiplets, including their flux
sectors, as well as the hatted fields.
The coupling $U_i\log\Phi_i$ and the mixed D-term
cannot be neglected in these integrations.
As we illustrate in Appendix~\ref{app:vortex},
integrating over another variable may change the
dominant saddle.

\subsection{FI deformations and polynomial deformations}

In the parameter map proposed by Morrison and
Plesser~\cite[eq.~(9)]{MP}, the FI parameters of one GLSM
correspond to the coefficients of the
superpotential of the other GLSM. It is useful to
study this correspondence in terms of the operators
in the parent theory.

Since $a_i=-L^2d_i$ and $\kappa_i=Ld_i$, varying
$\log d_i$ at fixed $L$ amounts to varying both
$\log a_i$ and $\log\kappa_i$ by the same amount.
Thus, the operator associated with this variation is
\begin{equation}
 \mathcal O_i=\Sigma_{H_i}+\Sigma_{K_i},\qquad
 o_i=\mathcal O_i|_0=x_i+y_i,
 \label{eq:O}
\end{equation}
where $x_i=\Sigma_{H_i}|_0$ and $y_i=\Sigma_{K_i}|_0$.
We evaluate the insertion at the original values of
the couplings.

Let $Q^A$ be a basis of the charge lattice
$\ker_{\mathbb Z}P$. In the GLSM obtained from the
A-side calculation, we have
\[
    \Sigma_{H_i}=\sum_AQ_i^A\Sigma_A,\qquad
    x_i=\sum_AQ_i^A\sigma_A,
\]
where $\sigma_A=\Sigma_A|_0$. As discussed in
Section~\ref{sec:forward}, the additional massive
pairs do not give a field-dependent one-loop correction
to the ordinary FI parameters near the $S$-Higgs vacuum.
We therefore expect that, after subtracting vacuum
constants, the marginal part of $o_i$ is
$\sum_AQ_i^A[\sigma_A]$, up to the normalization
of the FI terms. Here we assume that this identification
continues to hold after the remaining gauge integrations.

On the B side, we evaluate the same operator on
the solution \eqref{eq:saddle}. Since $o_i|_*=s_{i,*}$,
we have
\begin{equation}
 \left.o_i\right|_*-\kappa_i
 =\frac{d_ib_i}{1-b_i/L}
 =d_ib_i+O(b_i^2).
 \label{eq:BBoperator}
\end{equation}
The first term is precisely the term obtained by
varying $w_{\rm BB}$ with respect to $\log d_i$.
Thus, for a variation $\eta_i=\delta\log d_i$,
the proposed operator correspondence reads
\begin{align}
 \mathcal O(\eta)
 &=\sum_i\eta_i(\Sigma_{H_i}+\Sigma_{K_i}),
 \qquad o(\eta)=\mathcal O(\eta)|_0,
 \nonumber\\
 [o(\eta)]_{A,\rm marg}
 &=\sum_A\left(\sum_iQ_i^A\eta_i\right)[\sigma_A],
 \nonumber\\
 [o(\eta)]_{B,\rm poly}
 &=\left[\sum_i\eta_i d_ib_i\right].
 \label{eq:generalFImap}
\end{align}
For an LG orbifold with an isolated critical point,
the last expression denotes a class in the invariant
Jacobi ring. In a geometric or hybrid phase, we need
to consider the polynomial deformation globally,
including the gauge quotient and the redefinitions
of the equations defining the complete intersection.

Some of the variations of $d_i$ can be absorbed
into field redefinitions. In fact, from the definition
of $b_i$, we have
\begin{equation}
 \sum_iP_{ai}d_ib_i
 =\widehat\phi_a
   \partial_{\widehat\phi_a}w_{\rm BB}.
 \label{eq:generalEulerrelations}
\end{equation}
Thus, for $\eta\in\operatorname{im}P^T$, the variation
on the B side can be obtained by rescaling the hatted
fields. On the A side, we have $Q^A\cdot\eta=0$ for
every $A$. We therefore obtain the same quotient
$\mathbb C^n/\operatorname{im}P^T$ on both sides.
This does not necessarily give the number of physical
marginal operators. There can be more general field
redefinitions on the B side, and some FI deformations
of the ambient GLSM may be trivial in the compact
theory on the A side.

One should be careful when using the Jacobi ring
description in the geometric phase. For example,
in the local coordinates used in
Section~\ref{subsec:BBgeometriccritical}, the
superpotential is $\sum_\beta\widehat p_\beta z_\beta$.
A deformation of the form
$\widehat p_\beta\,\delta\widehat G_\beta$ can always
be removed locally by redefining $z_\beta$. However,
the field redefinitions on different coordinate patches may
not agree on their overlaps. Thus, such an operator
can be trivial in the local Jacobi sheaf and still
give a nontrivial complex-structure deformation.

Of course, not all complex-structure and K\"ahler
deformations are described by the parameters of
the toric GLSMs. There can also be deformations
associated with twisted-sector operators in an
orbifold description. It would be interesting
to see how the mirror correspondence for these
operators arises from the parent theory.

\subsection{Central charge}

In the geometric phase, the compact critical component
$\widehat Y$ obtained from the B-side calculation is
the expected BB mirror of $Y$. In an LG or hybrid phase,
we consider the corresponding compact sector.
The rational superpotential may also have other
critical points. The relation between these critical
points and the vacua of the parent theory is not
completely determined by the above calculations.
We will come back to this point in the quintic example
in Section~\ref{subsec:quinticexample}.

The central charges of the two compact GLSMs can be
computed from their geometric R-charges~\cite[Sec.~2]{MP}.
Using $\sum_i\mu_i=\sum_a\nu_a=r$ and $\rho=D+r$, we find
\begin{equation}
 \frac{c_A}{3}
 =\sum_{i=1}^n(1-2\mu_i)-(n-\rho)
 =\rho-2r=D-r
 =\sum_{a=1}^u(1-2\nu_a)-(u-\rho)
 =\frac{c_B}{3}.
 \label{eq:BBcentralcharge}
\end{equation}

Suppose that the selected IR theory has total central
charge $c_A$, is unitary, and contains the compact B-side
theory as a conformal factor. Under these assumptions, any
independent residual factor has $c=c_A-c_B=0$, by additivity
of the central charge. For a vacuum of unit norm, the
Virasoro commutator $[L_2,L_{-2}]=4L_0+c/2$ gives
\[
    \|L_{-2}|0\rangle\|^2=\frac{c}{2},
\]
so the stress tensor of the additional factor would be null.

Equation~\eqref{eq:BBcentralcharge} gives the central charges
of the two compact endpoints. Applying this argument to the
graded parent would also require its physical IR R-current,
unitarity, and the above factorization. For example, consider the light
fields $\Phi,\Theta_1,\Theta_2$ and the $H$ gauge
multiplet near $B=0$. Their contribution to the trial
R-anomaly is
\[
    (1-R_\Phi)-2(1-r_\Theta)-1=0,
    \qquad r_\Theta=1+R_\Phi/2.
\]
This equality does not depend on whether the fields
become massive in the IR. Thus, the vanishing of
this contribution does not necessarily mean that
there are no additional massless fields.
Also, the central charge calculation cannot distinguish
different finite gauge projections or exclude an
additional topological sector.

\subsection{Redundant deformations}

Morrison and Plesser use field rescalings to reach the two
parameter regions of their parent theory~\cite[Sec.~5]{MP}. We would
like to know whether this is still the case for our
extended parent theory.

From \eqref{eq:BBmodulimatching}, we know that changing
$L$ at fixed $c_a$ and $d_i$ does not change the
algebraic parameters of the two GLSMs. Furthermore,
in the geometric phase, the $L$ dependence of the
superpotential can be absorbed locally into a
redefinition of the normal coordinates. However,
such a field redefinition also changes the kinetic
terms. As we will see in the quintic example, some
of the higher-order terms can even be nontrivial in
the LG Jacobi ring, although they are irrelevant
deformations.

Since $a_i=-L^2d_i$ and $\kappa_i=Ld_i$, the operator
associated with varying $\log L$ at fixed $c_a$
and $d_i$ is
\[
    \mathcal I_L=\sum_i
       (2\Sigma_{H_i}+\Sigma_{K_i}),\qquad
    i_L=\mathcal I_L|_0=\sum_i(2x_i+y_i).
\]
Evaluating this operator on the B side, we have
\begin{equation}
 (i_L)_{B}-\sum_i\kappa_i
 =\partial_{\log L}w_L
 =-\frac1L\sum_i
   \frac{d_i b_i^2}{(1-b_i/L)^2}.
 \label{eq:Lresponse}
\end{equation}
Note that the right-hand side belongs to $\mathcal I^2$,
since each $b_i$ is linear in a $\widehat{\mathcal P}$-field.
Thus, in the geometric phase, changing $L$ does not
change the complex structure of $\widehat Y$ to first
order. Similarly, for an LG or hybrid model with the
IR grading discussed in Section~\ref{sec:fixed},
there is no marginal term in \eqref{eq:Lresponse}.

On the A side, from the Calabi--Yau condition, we have
\[
    2\sum_i x_i
    =2\sum_A\sigma_A\sum_iQ_i^A=0.
\]
The remaining contribution comes from the $K$ gauge
multiplets, which may give a vacuum constant in
the local Higgs calculation.

Thus the local Higgs calculation gives no physical FI
term from $2\sum_i x_i$, and the reduced B-side
superpotential gives no marginal term under the stated
grading. The projection of $\sum_i y_i$ after the full
gauge integration is still needed to establish quantum
redundancy of $L$.
In this paper, we keep $L$ finite and fixed when
taking the IR limit $E\to0$.

\subsection{A conjecture}
\label{subsec:irconjecture}

Let us consider a Batyrev--Borisov mirror pair such that both GLSMs
flow to smooth compact SCFTs. We define the gauge groups $G$ and
$\widehat G$ from the toric data, and the parameters are matched as
discussed above.

At fixed finite $L$ and fixed microscopic couplings, we take $E\to0$.
We conjecture that the source-free
$S$-Higgs sector of our graded parent theory and the compact sector
obtained from the B-side calculation are mirror descriptions of the same
IR SCFT. Under this correspondence, the ordinary chiral sector on one
side maps to the twisted chiral sector on the other.

In the geometric phase, the compact sector on the B side is associated
with $\widehat Y$. In a hybrid phase, one should keep the degrees of
freedom of the base as well, while in an orbifold phase one should
include the corresponding orbifold projection and the twisted sectors.

One may ask what happens to the fields that we integrated out in the
B-side calculation. Near $\widehat Y$, the Coulomb masses of $\Phi_i$
and of the two Grassmann-odd fields $\Theta_{i1}$ and $\Theta_{i2}$ tend
to zero. A related issue can be seen in the forward reduction
of the original MP construction~\cite[Sec.~5]{MP}.
On the regular branch, the hatted matter has Coulomb mass
$\widehat\sigma_a=c_a(A_a|_0)$ in our conventions. Each $A_a$
contains a $\mathcal P_\alpha$-field with scalar $p_\alpha$,
so this mass formula tends to zero as $p_\alpha$ tends to
zero on the geometric branch.

In our extended theory, we expect that the fields which become light
near $\widehat Y$ do not give an additional massless sector. To check
this, one needs to keep these fields and study their dynamics with the
$H$, $K$ and $\widehat H$ gauge multiplets all dynamical. We should also
recover the finite gauge projections and the correspondence between
physical operators from this calculation. The conjecture allows for
other massive vacua in the parent theory.

The construction applies to general Batyrev--Borisov nef partitions. In
the geometric phase, our local argument assumes that the defining
equations intersect transversely, so that the complete intersection is
smooth. At a singular point, the differentials of the defining equations
no longer give independent normal coordinates, and one has to study the
light fields directly in the GLSM.

\section{Examples}
\label{sec:examples}

\subsection{The quintic}
\label{subsec:quinticexample}

\subsubsection{Monomials and the original GLSM}

Consider the GLSM for the six-monomial family of
quintics~\cite{Witten,GP}, with five chiral superfields
$\mathcal U_j$ of charge one and a chiral superfield
$\mathcal P$ of charge $-5$ under a $U(1)$ gauge group.
Their lowest components are denoted by $u_j$ and $p$,
respectively. We use the following ordering of the
superfields:
\[
 \Phi=(\mathcal P,\mathcal U_1,\ldots,\mathcal U_5),
 \qquad
 \widehat\Phi=(\widehat{\mathcal P},
 \widehat{\mathcal U}_1,\ldots,\widehat{\mathcal U}_5).
\]
The corresponding matrix $P$ is
\begin{equation}
 P=\begin{pmatrix}
 1&\mathbf 1^T\\
 \mathbf 1&5I_5
 \end{pmatrix},
 \qquad
 \ell=(-5,1,1,1,1,1)^T,\qquad
 P\ell=P^T\ell=0.
 \label{eq:P}
\end{equation}
The monomials on the hatted side are therefore
\begin{equation}
 B_0=\widehat{\mathcal P}\prod_{j=1}^5
      \widehat{\mathcal U}_j,\qquad
 B_j=\widehat{\mathcal P}\,\widehat{\mathcal U}_j^5.
 \label{eq:B}
\end{equation}

In the $S$-Higgs vacuum, after integrating out the
additional massive fields, we can apply the calculation
of Morrison and Plesser. The gauge group on the A side
is $U(1)$ with charge vector $\ell$, and the resulting
superpotential is
\begin{equation}
 W_A=\mathcal P\left(c_0\prod_{j=1}^5\mathcal U_j
             +\sum_{j=1}^5c_j\mathcal U_j^5\right).
 \label{eq:WA}
\end{equation}
Denote the polynomial in parentheses by $G_5$.
The D-term equation reads
\[
 \sum_j|u_j|^2-5|p|^2=\xi_A.
\]
For a positive FI parameter $\xi_A$, not all $u_j$
can vanish. Assuming that $G_5=0$ defines a smooth
hypersurface, the F-term equations
\[
 G_5(u)=0,\qquad p\,dG_5(u)=0
\]
imply $p=0$. For a negative FI parameter, $p$ cannot
vanish, and we can use its expectation value to Higgs
the $U(1)$ gauge symmetry to $\mathbb Z_5$. The remaining
theory is the LG orbifold with superpotential
$G_5(\mathcal U)$. These are the two phases of the
quintic GLSM discussed in~\cite{Witten}.

\subsubsection{The quintic critical point at finite
\texorpdfstring{$L$}{L}}

Let us consider the parameter choice \eqref{eq:ray}
for the quintic. From the charge vector $\ell$, the
invariant combination of the coefficients is
\[
 z(d)=d_0^{-5}\prod_{j=1}^5d_j.
\]
We have $z(a)=z(d)$, while the hatted FI parameter is
chosen according to the coefficients $c_a$ of the
original quintic.

The B-side calculation gives the superpotential
\begin{equation}
 W_L(B)=\sum_{i=0}^5\frac{d_iB_i}{1-B_i/L},
 \label{eq:WL}
\end{equation}
where we have subtracted the six field-independent
constants. In the LG phase, we set
$\widehat{\mathcal P}=1$ and denote
$\widehat{\mathcal U}_j$ by $\mathcal X_j$.
The terms linear in $B_i$ then give
\begin{equation}
 W_5(\mathcal X)=d_0\prod_{j=1}^5\mathcal X_j
                 +\sum_{j=1}^5d_j\mathcal X_j^5.
 \label{eq:W5superfield}
\end{equation}
In terms of the lowest components
$X_j=\mathcal X_j|_0$, the expansion of $w_L=W_L|_0$
around the origin reads
\begin{align}
 w_L&=w_5+\frac1L\sum_i d_i b_i^2+O(X^{15}),
 \nonumber\\
 w_5&=d_0\prod_jX_j+\sum_jd_jX_j^5,
 \qquad b_0=\prod_jX_j,\quad b_j=X_j^5.
 \label{eq:W5}
\end{align}

Away from the discriminant locus of the quintic,
the five quartic polynomials $\partial_jw_5$ form
a regular sequence, with their common zero at the
origin. They are also the lowest-degree terms of
$\partial_jw_L$. Since these lowest-degree terms form a regular sequence,
the associated graded Jacobi algebra of $w_L$ is
$\Jac(w_5)$. Its Hilbert series is $(1+z+z^2+z^3)^5$,
so its dimension before the orbifold projection is
$4^5=1024$ for every finite $L\ne0$. This is the local
Milnor algebra of a semiquasihomogeneous singularity
with principal part $w_5$~\cite{GHP}. The multiplication
in the Jacobi algebra of $w_L$ may, however, depend
on $L$.

At the IR fixed point associated with $W_5$, each
$X_j$ has weight $1/5$~\cite{IV}. Since all the $b_i$
are of degree five, the operator $b_i^n$ has weight
$n$ whenever it survives in the Jacobi ring.
Its F-term coupling has RG eigenvalue $1-n$.
Thus, we expect the higher-order terms in
\eqref{eq:W5} to be irrelevant, without taking
the limit $L\to\infty$.

\subsubsection{Higher-order terms}

The relation $d_i=-\kappa_i^2/a_i$ gives
\[
 z(d)=\frac{z(\kappa)^2}{z(a)}.
\]
With $\kappa_i=Ld_i$, we have already matched the
algebraic parameters at the microscopic scale.
It is interesting to see whether the difference
between $w_L$ and $w_5$ is simply due to a field
redefinition.

Write $\beta_i=d_ib_i$. From the derivatives of
$w_5$, we have the relations
\[
 5\beta_j+\beta_0=0
\]
in the Jacobi ring. The degree-ten term in
\eqref{eq:W5} therefore gives
\begin{equation}
 \left[\frac1L\sum_i d_i b_i^2\right]
 =\frac{d_0}{L}
  \left(1+\frac{d_0}{25}\sum_{j=1}^5d_j^{-1}\right)
  [b_0^2].
 \label{eq:irrelevantclass}
\end{equation}
To see whether $[b_0^2]$ is nontrivial, consider
the Fermat limit of $w_5$. The Jacobi ring in this
limit is
\[
 \C[X_1,\ldots,X_5]/(X_1^4,\ldots,X_5^4),
\]
in which $b_0^2=\prod_jX_j^2$ is clearly nonzero.
This class remains nonzero in an open neighborhood
of the Fermat point. Hence, for generic nonzero
$d_0$ in this neighborhood, the degree-ten term
in \eqref{eq:W5} cannot be absorbed into an
infinitesimal holomorphic field redefinition.

However, the weight of this operator is two.
More generally, if a chiral primary has weight $n$,
its integrated F-term has dimension $n+1$. Thus,
to first order, the corresponding dimensionless
coupling at energy $E$ is related to that at
energy $E_0$ by
\[
 g_n(E)=g_n(E_0)
 \left(\frac{E}{E_0}\right)^{n-1}.
\]
Although the degree-ten term gives a nontrivial
deformation at the matching scale, it is irrelevant
in the reduced LG theory. For the parent theory,
we still need to take into account the remaining
gauge dynamics.

\subsubsection{The orbifold and the other physical modulus}

We take the integral factorization of $P$ for which
the ordinary gauge group is connected. Denote a
phase transformation of the hatted fields by
$(g_0,g_1,\ldots,g_5)$. For such a transformation
to preserve all $B_i$, we require
\[
 g_0g_j^5=1,\qquad g_0\prod_jg_j=1.
\]
In the LG phase, where $\widehat p=1$, we have
$g_0=1$. The remaining group is
\begin{equation}
 \Gamma=\{(\omega_1,\ldots,\omega_5)\in\mu_5^5:
                          \ \prod_j\omega_j=1\},
 \qquad
 X_j\longmapsto\omega_jX_j,
 \label{eq:group}
\end{equation}
where $\mu_5$ denotes the group of fifth roots of unity.
Dividing $\Gamma$ by the diagonal $\mu_5$, which is
part of the projective identification in the geometric
phase, gives the usual $(\Z/5)^3$ action in~\cite{GP}.
The integral factorization and the gauge groups are
discussed in Appendix~\ref{app:integral}.

Here we have identified the group from the integral
data of the two GLSMs. To obtain the same orbifold
from the parent theory, we need to perform the sum
over the hatted gauge bundles.

The LG orbifold $(W_5(\mathcal X),\Gamma)$ describes
the mirror quintic at the LG point of its K\"ahler
moduli space. Varying the hatted FI parameter away
from this point turns on the marginal deformation from
the diagonal twisted sector~\cite[eqs.~(3.44)--(3.46)]{Witten}.
We keep this
modulus fixed in the following calculation.

\subsubsection{The marginal operator in the quintic example}

For the quintic, we can make the operator correspondence
in \eqref{eq:generalFImap} more explicit. The lowest
component of the insertion
$\mathcal O_i=\Sigma_{H_i}+\Sigma_{K_i}$ on the B side,
after subtracting the vacuum constant, is
\[
 \left.o_i\right|_*-\kappa_i
 =\frac{d_ib_i}{1-b_i/L}.
\]
We are interested in its marginal part $d_ib_i$.

Consider first the five coordinate columns. The
logarithmic derivatives of $w_5$ give
\begin{equation}
 X_j\partial_jw_5=5d_jX_j^5+d_0\prod_kX_k,
 \qquad
 [d_jX_j^5]=-\frac{d_0}{5}
                  \Big[\prod_kX_k\Big].
 \label{eq:Jac}
\end{equation}
There are six invariant monomials of degree five
under $\Gamma$, namely $b_0,\ldots,b_5$. Indeed, invariance
under opposite fifth-root rotations of any pair of fields
requires their exponents to agree modulo five.
Among the degree-five Jacobi relations
$X_k\partial_jw_5$, only those with $k=j$ are invariant.
For $d_j\ne0$, these five relations are independent.
Hence, there is one marginal operator in the
invariant polynomial sector, which we can take to
be $[b_0]$.

In the reduced A-side GLSM, each of the five coordinate
fields has charge one, so each coordinate column projects
onto the same FI operator $[\sigma]$. If this projection
persists in the joint gauge integral of the parent, we
expect the correspondence
\begin{equation}
 [\sigma]\ \longleftrightarrow\
 -\frac{d_0}{5}\Big[\prod_kX_k\Big].
 \label{eq:map}
\end{equation}
For the $\mathcal P$-field column, we obtain $-5[\sigma]$
on the A side and $d_0[b_0]$ on the B side, which
gives the same correspondence.

For a general variation $\eta_i=\delta\log d_i$
with $L$ fixed, the operator in the parent theory is
\[
 \mathcal O(\eta)=\sum_i\eta_i\mathcal O_i,\qquad
 o(\eta)=\mathcal O(\eta)|_0
         =\sum_i\eta_i(x_i+y_i).
\]
Using \eqref{eq:Jac}, we obtain the marginal
projections on the two sides,
\begin{align}
 [o(\eta)]_{A,\rm marg}
    &=(\ell\cdot\eta)[\sigma],\\
 [o(\eta)]_{B,\rm marg}
    &=-\frac{d_0}{5}(\ell\cdot\eta)[b_0].
 \label{eq:physicaldirection}
\end{align}
We see that if $\delta\log z(d)=\ell\cdot\eta=0$,
both marginal projections vanish. Of the six variations
$\delta\log d_i$ at fixed $L$, only the combination
$\delta\log z(d)$ couples to the invariant polynomial
marginal operator in this family.

The normalization of these operators in correlation
functions has not been determined. Also, the above
calculation does not give the mirror map for operators
in the twisted sectors.

\subsubsection{The interacting critical point in the LG phase}

The superpotential $w_L$ can have critical points
other than the origin. In Appendix~\ref{app:geometry},
we show that, for generic coefficients, all the other
finite critical points in the $X_j$ variables away from
the poles of $w_L$ are Morse critical points.
The local Jacobi algebra at each such point is one-dimensional.
Moreover, all these points lie on the dense torus, where
$\Gamma$ acts freely. Thus no twisted-sector state is
localized at any of them.

This is of interest for the operator correspondence
discussed above. If the FI deformation of the
original GLSM survives the remaining gauge
integrations, we should obtain a marginal operator
on the B side. Among the finite critical points
of $w_L$ on this branch, only the origin can account
for this operator. The vortex calculation in
\eqref{eq:bridge} relates the $S$-Higgs vacuum to
the auxiliary Coulomb solution used here, but the
$H$ gauge multiplets were kept fixed in that
calculation.

The central charge of the quintic is $c=9$.
The five LG fields on the B side give~\cite{IV,LVW}
\[
 c=3\sum_{j=1}^5\left(1-\frac25\right)=9,
\]
and this value is unchanged by the orbifold.
The question of identifying the stress tensor
of the parent theory was discussed in
Section~\ref{sec:conjecture}.

There is another interesting aspect of the quintic
example. With the charge vector
$\ell=(-5,1,1,1,1,1)$, the condition
\eqref{eq:signgrading} becomes
\[
 -5\varepsilon_0+\sum_{j=1}^5\varepsilon_j=0.
\]
Since $\varepsilon_i=\pm1$, this equation implies
\[
 \varepsilon_0=\varepsilon_1=\cdots=\varepsilon_5.
\]
Thus, we cannot change the sign of the $\mathcal P$-field
column alone if we require a common grading.
This is a particular property of the quintic
charge vector. For a general matrix $P$, there
can be other solutions to \eqref{eq:signgrading}.

\subsection{A codimension-two example: two cubics in projective space}
\label{subsec:twocubics}

Consider the complete intersection of two cubics in
$\mathbb P^5$. The GLSM has two chiral superfields
$\mathcal P_\alpha$ and six chiral superfields
$\mathcal X_j$, with
\begin{equation}
 \ell=(-3,-3,1,1,1,1,1,1)^T,\qquad
 W_A=\mathcal P_1G_1(\mathcal X)
      +\mathcal P_2G_2(\mathcal X).
 \label{eq:cubiccharges}
\end{equation}
Here we order the fields as
$(\mathcal P_1,\mathcal P_2,\mathcal X_1,\ldots,
\mathcal X_6)$.

Take the nef partition $I_1=\{1,2,3\}$ and
$I_2=\{4,5,6\}$. A cubic polynomial in six variables
has 56 monomials. Thus, for general $G_1$ and $G_2$,
the matrix $P$ has 112 rows and eight columns.
The row associated with the monomial
$\mathcal P_\alpha\prod_j\mathcal X_j^{k_j}$ is
\begin{equation}
 P_{(\alpha,k),\bullet}
 =(\delta_{\alpha1},\delta_{\alpha2},k_1,\ldots,k_6),
 \qquad
 \alpha=1,2,\qquad k_j\geq0,\quad\sum_jk_j=3.
 \label{eq:cubicP}
\end{equation}
For a vector in $\ker P$, subtracting the row equations
for $\mathcal P_\alpha\mathcal X_j^3$ and
$\mathcal P_\alpha\mathcal X_k^3$ shows that all six
coordinate entries are equal. Each of the first two entries
is then minus three times this common value. Hence
$\ker_{\Z}P=\Z\ell$ and $\operatorname{rank}P=7$.

The vector $\boldsymbol\mu$ in \eqref{eq:BBgrading} is
$(1,1,0,\ldots,0)^T$. For $\boldsymbol\nu$, we choose its entries
to be one on the rows associated with
\[
 \mathcal P_1\mathcal X_1\mathcal X_2\mathcal X_3,
 \qquad
 \mathcal P_2\mathcal X_4\mathcal X_5\mathcal X_6,
\]
and zero on the other rows. It is easy to see that
\[
 P\boldsymbol\mu=\mathbf1,\qquad P^T\boldsymbol\nu=\mathbf1,
 \qquad \sum_i\mu_i=\sum_a\nu_a=2.
\]
Thus, $\rho-2r=7-4=3$, as expected for this
Calabi--Yau threefold.

To give the factorization $P=\mathsf S\mathsf T$, denote the rays
of $\mathbb P^5$ by $v_j=e_j$ for $j=1,\ldots,5$
and $v_6=-\sum_{j=1}^5e_j$. The columns of $\mathsf T$
are given by
\begin{equation}
 \bar n_{p_\alpha}=(0,e_\alpha),\qquad
 \bar n_{x_j}=(v_j,e_{\beta(j)}),\qquad
 \beta(j)=
 \begin{cases}
 1,&j\in I_1,\\
 2,&j\in I_2,
 \end{cases}
 \label{eq:cubicT}
\end{equation}
and the rows of $\mathsf S$ are given by
\begin{equation}
 \bar m_{\alpha,k}
  =\bigl(k_1-\mathbf1_{1\in I_\alpha},\ldots,
         k_5-\mathbf1_{5\in I_\alpha};e_\alpha\bigr).
 \label{eq:cubicS}
\end{equation}
For $j\leq5$, the last two entries in the pairing with
$\bar n_{x_j}$ cancel the subtraction
$\mathbf1_{j\in I_\alpha}$. For $j=6$, the relation
$v_6=-\sum_{j=1}^5e_j$, together with $|I_\alpha|=3$,
gives $3-\sum_{j=1}^5 k_j=k_6$. This reproduces
\eqref{eq:cubicP}.

The two rows $(0,e_1)$ and $(0,e_2)$ of $\mathsf S$
correspond to the fields
$\widehat{\mathcal P}_1$ and
$\widehat{\mathcal P}_2$ of the mirror GLSM.
From \eqref{eq:cubicP}, the monomials $B_i$
associated with $p_1,x_1,x_2,x_3$ contain
$\widehat{\mathcal P}_1$, and those associated
with $p_2,x_4,x_5,x_6$ contain
$\widehat{\mathcal P}_2$. The mirror superpotential
therefore takes the form
\[
 W_{\rm BB}
 =\widehat{\mathcal P}_1\widehat G_1(\widehat{\mathcal U})
 +\widehat{\mathcal P}_2\widehat G_2(\widehat{\mathcal U}).
\]

In this example we have eight auxiliary systems
in the parent theory. Performing the same calculation
for each of them, with the parameter choice
\eqref{eq:ray}, we obtain
\begin{equation}
 W_L(B)=\sum_{i=1}^8\frac{d_iB_i}{1-B_i/L},
 \qquad
 z(d)=\frac{\prod_{j=1}^6d_{x_j}}
             {d_{p_1}^3d_{p_2}^3}.
 \label{eq:cubicthreshold}
\end{equation}
The terms linear in $B_i$ give $W_{\rm BB}$.
For a smooth mirror complete intersection, the
higher-order terms can be absorbed locally into
the two normal coordinates, as discussed in
Section~\ref{subsec:BBgeometriccritical}.

The rank of the gauge group on the hatted side
is $112-7=105$ in this GLSM presentation.
Of course, not all these FI parameters necessarily
give independent marginal deformations of the
compact theory.

\subsubsection{The hybrid chamber and the compact period}

An interesting difference between this example and
the quintic is the phase with negative FI parameter.
The D-term equation now requires that $p_1$ and $p_2$
cannot vanish simultaneously. The quotient of these
two fields gives $\mathbb P(3,3)$, with coarse space
$\mathbb P^1$, and there is a residual $\mathbb Z_3$
gauge symmetry acting on the six fields $x_j$.
The superpotential at a given point of the base is
\[
 w_A=p_1G_1(x)+p_2G_2(x).
\]
Thus, we have a hybrid model rather than a pure
LG orbifold~\cite{EKhybrid,BMPHybrid}.

In this phase, by mixing the R-current with the
gauge current, we can assign R-charge zero to
$\mathcal P_\alpha$ and R-charge $2/3$ to
$\mathcal X_j$. The central charge of the hybrid
model is
\begin{equation}
 \widehat c=1+6(1-2/3)=3.
 \label{eq:cubicHybridCentral}
\end{equation}
The first term comes from the base $\mathbb P^1$,
which is absent in the LG description of the
quintic.

Recall also the Hori--Vafa mirror of this model
discussed in Section~\ref{sec:HV}. On the dense
torus, it is given by the equations
\[
 \prod_{j=1}^6\mathsf U_j=e^{-t},\qquad
 1+\mathsf U_1+\mathsf U_2+\mathsf U_3=0,\qquad
 1+\mathsf U_4+\mathsf U_5+\mathsf U_6=0.
\]
In passing from the noncompact model to the
compact model, each $\mathcal P_\alpha$ has gauge charge
$-3$ and gives an insertion $3\partial_t$. The relation
between the periods is therefore
\[
 \Pi_{\rm c}\propto(3\partial_t)^2\Pi_{\rm nc}.
\]
This is the analogue of the single insertion
we discussed for the quintic in
Section~\ref{subsec:quinticmeasure}.

\section{Discussion}
\label{sec:discussion}

The original motivation of this work was to understand
whether the sign issue in the Morrison--Plesser mirror
construction could be resolved by using the right
fundamental variables on the mirror side. As we have
seen from the Hori--Vafa construction, two models
with the same superpotential can have different
integration measures. In the quintic example, however,
the local calculation with the flat reference measure in
Appendix~\ref{app:measure} gives a constant multiple of
the logarithmic volume form. The compact period requires
one additional insertion. It would still be interesting to compute the
full period directly from their parent theory.

In this paper, we have studied a different way of
resolving the sign issue by introducing Grassmann-odd
matter fields into the MP parent theory. Two such
fields can reverse the sign of the one-loop correction
from an ordinary chiral field. To make these additional
fields massive on the A side, we introduced their
partners and an auxiliary gauge group. The partners
also contribute to the effective twisted superpotential
on the B side. Instead of obtaining a term proportional
to $1/B$, we obtain
\[
 F(B)=\frac{\kappa a}{a+\kappa B}.
\]
The interesting point here is that the leading
field-dependent term is proportional to $B$, even
though we have not changed the logarithmic coupling
in the original MP action.

Since this calculation only involves the fields
associated with one column of $P$, it can be applied
to general Batyrev--Borisov mirror pairs. We have also
seen that, for a smooth complete intersection, the
higher powers of $B_i$ do not change the compact
critical component of the reduced theory. There
is no need to take $L$ to infinity in this calculation.

The vortex calculation and the FI insertions give two
checks on this proposal. At fixed ordinary gauge background,
the vortex sum selects the larger auxiliary Coulomb root in
the real domain of \eqref{eq:bridge}. At fixed $L$, the
FI insertions reproduce the degree-five Jacobi classes in
the quintic example. The joint gauge integral is still needed
to identify the compact sector. In particular, the Coulomb
masses of $\Phi_i$ and $\Theta_{iI}$ on the
regular B-side solution tend to zero with the corresponding
hatted monomial. In the forward MP calculation, the hatted mass is
$\widehat\sigma_a=c_a(A_a|_0)$, which vanishes on the smooth
geometric critical locus. The local thresholds alone do
not determine whether additional normalizable states survive.
To answer this question, one needs to study the
dynamics of the parent theory with both the ordinary
and hatted gauge fields dynamical. It would be
particularly interesting to compute correlation
functions of the FI operators in this theory,
including the sum over the gauge bundles.

There is also a question concerning the redundant
deformations used by Morrison and Plesser. Although
we have found a family of parameters labelled by $L$
for which the algebraic parameters of the two GLSMs
are the same, we do not know whether these deformations
are redundant in the full quantum theory.

In view of the above calculations, we conjecture
that our extended parent theory can give rise to
the two GLSMs associated with a Batyrev--Borisov
mirror pair by integrating out different sets of
fields, and that the corresponding compact sectors
flow to the same IR SCFT under mirror symmetry
at fixed microscopic couplings.

\section*{Acknowledgements}
We thank Ilarion Melnikov, Ronen Plesser, Mauricio Romo
and Eric Sharpe for bringing this problem to the author's
attention at different stages. In particular, we thank
Ilarion Melnikov for sharing his insights with us.
WG is supported by the National Natural Science Foundation
of China (Grant No.~12575077).  

\section*{AI disclosure}

The main ideas of this paper are the author's own. However, AI assisted the author in several aspects of this work. The author first learned of Morrison and Plesser's work during his Ph.D. studies but did not fully understand the details of their construction at the time. AI helped the author revisit their paper and better understand some of these details. In particular, AI assisted the author in carrying out the computations presented in this paper, which were subsequently checked by the author. AI also assisted in the writing and proofreading of this paper. The author takes full responsibility for the results and conclusions presented here.

\appendix

\section{The quintic unit-state insertion in the original MP model}
\label{app:measure}

We would like to see whether assigning R-charge two to the
$\mathcal P$-field in the MP parent theory gives the additional factor needed
for the compact period. For this purpose, we compare the matter
determinants at two different R-charges, keeping the Coulomb mass
and the gauge flux fixed.

For a chiral superfield of gauge charge one and vector R-charge $R_i$,
the corresponding Dolbeault complex after the A-twist is that of
$\mathcal L_i\otimes K^{R_i/2}$, where
$\deg\mathcal L_i=\mathfrak m_i$. On a genus-$g$ Riemann surface,
the Riemann--Roch formula gives~\cite[eq.~(2.38)]{CMP}
\begin{equation}
 \chi_i=\mathfrak m_i+(R_i-1)(g-1),\qquad
 Z_i^{\rm mat}=\sigma_i^{-\chi_i}.
\end{equation}
Thus, on the sphere, changing the R-charge of $\mathcal P$ from zero
to two gives an additional factor of $\sigma_0^2$ in the matter
determinant. This is independent of the gauge flux.
For $(R_0,R_1,\ldots,R_5)=(2,0,\ldots,0)$, the curvature coupling
is given by
\[
 e^{2\pi i\Omega_\chi}
 =q_R\frac{\prod_i\sigma_i}{\sigma_0^2},
\]
where we have expressed the masses in a common unit and $q_R$
depends on the convention for the curvature counterterm~\cite{CMP}.

Consider the gauge slice $\widehat p=1$ on the dense torus, with
the reference measure
$\Omega_{\rm lin}=\bigwedge_{j=1}^5d\widehat u_j$.
Since the masses of the ordinary matter fields depend on the
hatted fields, we should also take this dependence into account
when integrating over the Coulomb scalars. The lowest-component
function we consider here is
\begin{equation}
 \widetilde w_{\rm ref}(\sigma,\widehat u)
 =\sum_i\sigma_i
 \bigl[\log(m_i(\widehat u)/\sigma_i)+1\bigr],
 \qquad
 m_i(\widehat u)=\frac{a_i}{b_i(\widehat u)},\qquad
 \widetilde w_-=\sum_i m_i.
\end{equation}
The equations for the $\sigma_i$ give $\sigma_i=m_i$.
Writing
\[
 D=\operatorname{diag}(\sigma_i),\qquad
 J_{ij}=\partial_{\widehat u_j}\log m_i,\qquad
 K_{jk}=\sum_i\sigma_i
 \partial_{\widehat u_j}\partial_{\widehat u_k}\log m_i,
\]
we obtain
\begin{align}
 \widetilde w_{\rm ref}''&=
 \begin{pmatrix}-D^{-1}&J\\J^T&K\end{pmatrix},
 &
 K+J^TDJ&=\partial_{\widehat u}^2\widetilde w_-.
\end{align}
The determinant of the six-dimensional block $-D^{-1}$
cancels $\prod_i\sigma_i$ in the curvature coupling, so that
\begin{equation}
 \left.e^{2\pi i\Omega_\chi}
       \det\widetilde w_{\rm ref}''\right|_{\sigma=m}
 =\frac{q_R}{m_0^2}
   \det\partial_{\widehat u}^2\widetilde w_-.
 \label{eq:massiverefschur}
\end{equation}

Recall that for an LG model with holomorphic volume form
$v(\widehat u)\Omega_{\rm lin}$, the handle operator at a
nondegenerate critical point is $\det w''/v(\widehat u)^2$.
This follows from the flat-volume result~\cite[eqs.~(2.9)--(2.10)]{GuLG}
by a local change of variables with Jacobian determinant $v$.
In this normalization, \eqref{eq:massiverefschur} corresponds
locally to the volume form $m_0\Omega_{\rm lin}$, up to a constant.
Using $m_0=a_0\delta_0$ and $\widehat u_j=Y_j^{-1}$, we have
\begin{equation}
 m_0\Omega_{\rm lin}
 =a_0\delta_0\,(-\delta_0^{-1})\Omega_{\log}
 =-a_0\Omega_{\log}.
\end{equation}
The Jacobian of the inversion therefore cancels the factor
of $\delta_0$ from the mass of the $\mathcal P$-field. The resulting measure is still logarithmic
in the $Y_j$ variables.

In the above calculation, it is important that we have fixed
$\widehat p=1$. Otherwise, $\ell$ is a null vector of the
Hessian in the six logarithmic coordinates.

There is a further subtlety. Consider the critical equations
of
\[
 \widetilde w_-=\sum_j a_jY_j^5+a_0\prod_jY_j
\]
on the dense torus. By multiplying the five logarithmic
critical equations, we obtain
\[
 a_0^5=-5^5\prod_j a_j.
\]
Thus, for generic smooth quintic coefficients, there is no
critical point on the torus. Even when this relation is
satisfied, the critical point is not isolated, since a common
rescaling of the $Y_j$ gives another solution.

The above calculation only compares local volume forms.
We have not determined the integration cycle of the MP period
or performed the remaining gauge integrations. In particular,
it remains to be seen whether the mixed gauge term and the
couplings $U_i\log\Phi_i$ give the additional insertion
required for the compact period.

\section{Local column checks at fixed \texorpdfstring{$L$}{L}}
\label{app:local}

\subsection{The auxiliary equations}

Consider the two Coulomb equations in \eqref{eq:saddle}.
After clearing the denominators, we write them as
\begin{equation}
 F_1=ax-b(x+y)^2,\qquad
 F_2=(x+y)^2-\kappa y.
 \label{eq:cleared}
\end{equation}
On the branch with $y\ne0$, setting $s=x+y$ gives
\[
 y=\frac{s^2}{\kappa},\qquad
 s=\frac{a\kappa}{a+\kappa b}.
\]
For a given $b$, there is a unique solution on this branch,
provided that $a+\kappa b\ne0$. At $b=0$, the solution is
$(x,y)=(0,\kappa)$, where the Jacobian of \eqref{eq:cleared}
is
\begin{equation}
 \frac{\partial(F_1,F_2)}{\partial(x,y)}
 =\begin{pmatrix}a&0\\2\kappa&\kappa\end{pmatrix}.
 \label{eq:clearedJ}
\end{equation}
Since its determinant is $a\kappa\ne0$, the solution extends
regularly to $b=0$. However, the mass of the fields
$\Phi,\Theta_1,\Theta_2$ is $x$, which vanishes at this point.
We therefore consider the following partial integration.

\subsection{Keeping the light \texorpdfstring{$H$}{H} multiplets}

We first solve the $K$ Coulomb equation at fixed $x$.
Choosing the solution for which $y=\kappa$ at $x=0$, we have
\begin{equation}
 s_+(x)=\frac{\kappa+\sqrt{\kappa(\kappa-4x)}}{2}
       =\kappa-x+O(x^2/\kappa),
 \qquad
 \left.\partial_y^2\widetilde w_{\rm eff}\right|_{0,\kappa}
 =-1/\kappa.
\end{equation}
In a neighborhood of $x=0$, neither $y$ nor $s$ vanishes.
The one-loop contribution from $\Phi,\Theta_1,\Theta_2$
depends only on $x$ and hence does not enter the equation
for $y$.

For $x\ne0$, we can include the contribution from these
three fields and define
\begin{equation}
 \widetilde w_{\rm red}(x;b)
 =\widetilde w_{\rm eff}(x,y_*(x);b),
 \qquad y_*(x)=s_+(x)-x.
\end{equation}
Taking a derivative with respect to $x$, we obtain
\begin{equation}
 \partial_x\widetilde w_{\rm red}
 =\log\frac{qx}{s_+(x)^2}
 =\log\frac{x}{\Lambda_H(b)}
  +\frac{2x}{\kappa}+O(x^2/\kappa^2),
 \qquad
 \Lambda_H(b)=\frac{\kappa^2}{q}=-db.
 \label{eq:Hscale}
\end{equation}
If we keep $\Phi,\Theta_1,\Theta_2$ in the Wilsonian theory,
we should of course subtract their one-loop contribution
$+\mathcal L(x)$ from $\widetilde w_{\rm red}$.

One interesting observation is that $\Lambda_H(b)$ goes to
zero as $b\to0$, even though $\kappa$ is finite and nonzero.
Thus, we cannot use the mass scale of the $K$ sector to
argue that the remaining $H$ sector is massive.

On the other hand, $\Phi,\Theta_1,\Theta_2$ all have charge
one under $U(1)_H$. There is therefore no nonconstant
gauge-invariant polynomial constructed solely from these
three fields. This does not exclude gauge-invariant operators
involving other fields, disorder operators, or operators
that are regular only on patches where a charged field is nonzero.

Although the reduced LG superpotential has the expected
isolated quintic critical point, we do not yet know how this
point is realized in terms of the microscopic fields of
the parent theory. The calculations in
Section~\ref{sec:conjecture} concern this question on
regular backgrounds.

\subsection{The local handle}

For the massive Coulomb solution, the handle operator in
the convention of~\cite{CMP} is
\[
 \cH=e^{2\pi i\Omega}\det\widetilde w_{\rm eff}''.
\]
From the charges of the matter fields, we obtain, up to a
background-dependent constant,
\[
 e^{2\pi i\Omega_{\rm loc}}
 =x^{-1-R_\Phi+2r_\Theta}y\,s^{2(1-r_\Theta)}.
\]
Using $\det\widetilde w_{\rm eff}''=-1/(xy)$ and
$r_\Theta=R(\Theta_I)=1+R_\Phi/2$, we have
\begin{equation}
 \cH_{\rm loc}
 =-s_*^{-R_\Phi}
 =-\kappa^{-R_\Phi}(1-b/L)^{R_\Phi}.
 \label{eq:handle}
\end{equation}

For the coordinate columns, $R_\Phi=0$ and hence
$\cH_{\rm loc}=-1$. For the $\mathcal P$-field column, $R_\Phi=2$,
and we have
\[
 \frac{\cH_{\rm loc}(b)}{\cH_{\rm loc}(0)}
 =(1-b/L)^2,
\]
where $\cH_{\rm loc}(0)$ denotes the continuation of the
massive result to $b=0$. This factor is regular and nonzero
at the origin. At $b=0$, however, we should also take into
account the contribution from the light fields discussed
above and their couplings to the remaining gauge multiplets.

\subsection{The relative flux calculation}

We can also obtain \eqref{eq:handle} from a flux calculation
with an insertion $x^u y^v$. We take $u,v$ to be nonnegative
integers and use the expansion $|x|\ll|y|$ when evaluating
the residues.

On a genus-$g$ Riemann surface, let $(m,k)\in\Z^2$ denote
the gauge fluxes of $H$ and $K$, respectively. In the absence
of flavor backgrounds, an ordinary chiral superfield of mass
$m_{\rm mat}$, gauge charges $Q$, and vector R-charge
$r_{\rm mat}$ has the one-loop determinant~\cite[eq.~(2.38)]{CMP}
\[
 Z_{\rm chiral}^{\rm 1-loop}=m_{\rm mat}^{-\chi},
 \qquad
 \chi=Q\cdot(m,k)+(r_{\rm mat}-1)(g-1).
\]
Here $\chi$ is the Riemann--Roch index of the twisted matter
fields. For a Grassmann-odd chiral superfield, reversing the
statistics gives the inverse determinant, as in the sphere
calculation of~\cite[eqs.~(15a) and (15b)]{GuZou}.

With $R(\Theta_I)=1+R_\Phi/2$ and $R(\Xi_I)=1-R_\Phi/2$,
the fields $\Phi,\Theta_1,\Theta_2$ give a factor of
$x^{m+g-1}$, while $S,T_1,T_2$ give $y^{k+g-1}$.
The two $\Xi_I$ fields contribute
$(x+y)^{-2(m+k)+R_\Phi(1-g)}$.
For each of the $g$ pairs of harmonic one-forms, the $H$ and
$K$ gaugino zero modes couple through the two-variable Coulomb
Hessian. Their integral gives one determinant per handle
in the fixed-flux formula~\cite[eq.~(2.37)]{CMP}. With
$s=x+y$, the Hessian in \eqref{eq:HKHess} gives
\[
 \det\widetilde w_{\rm eff}''
 =\frac1{xy}-\frac2s\left(\frac1x+\frac1y\right)
 =-\frac1{xy}.
\]
Multiplying by the matter determinants gives
\begin{equation}
 \begin{aligned}
 I_{m,k}^{(g)}
 &=x^{m+g-1}y^{k+g-1}s^{-2(m+k)+R_\Phi(1-g)}
   \left(-\frac1{xy}\right)^g\\
 &=(-1)^g x^{m-1}y^{k-1}(x+y)^{-2(m+k)+R_\Phi(1-g)}.
 \end{aligned}
 \label{eq:formalfluxlong}
\end{equation}
Here we have absorbed constant phases into the FI parameters
and used the same normalization for the neutral matter
determinant as in the previous calculation.

For a polynomial insertion $\mathcal O(x,y)$, we define the
genus-$g$ flux sum for the $H\times K$ sector by
\[
 \mathcal R_g(\mathcal O)
 :=\sum_{(m,k)\in\Z^2}q^m\kappa^k\,
 [x^{-1}y^{-1}]
 \bigl(\mathcal O(x,y)\,I_{m,k}^{(g)}(x,y)\bigr).
\]
The brackets extract the coefficient of $x^{-1}y^{-1}$ after
expansion in $x/y$ for $|x|<|y|$.
We now expand $x^u y^v I_{m,k}^{(g)}$ in powers of $x/y$.
Let $j$ denote the power of $x/y$ in the expansion of
$(x+y)^{-2(m+k)+R_\Phi(1-g)}$. A term can contribute to the
coefficient of $x^{-1}y^{-1}$ only if
\[
 m=-u-j,\qquad
 k=v+2u+j+R_\Phi(1-g),\qquad j\geq0.
\]
For these fluxes, the upper index of the binomial coefficient
is $-2(u+v)-R_\Phi(1-g)$, independent of $j$. Summing over
$j$ with the FI weights $q^m\kappa^k$, we obtain
\begin{align}
 \mathcal R_g(x^u y^v)
 &=(-1)^g q^{-u}\kappa^{v+2u+R_\Phi(1-g)}
   (1+\kappa/q)^{-2(u+v)-R_\Phi(1-g)}
 \nonumber\\
 &=(-1)^g s_*^{R_\Phi(1-g)}x_*^u y_*^v,
 \qquad |\kappa/q|<1,
 \label{eq:formaltrace}
\end{align}
where $s_*=x_*+y_*$. With the coefficient convention used
above and $\cH_{\rm loc}=-s_*^{-R_\Phi}$, this reads
$\mathcal R_g(x^uy^v)=-\cH_{\rm loc}^{\,g-1}x_*^uy_*^v$.
Reversing the overall residue orientation gives the usual
Bethe-vacuum normalization, including the unit trace at $g=1$.

We can compare this result with the vortex calculation in
Section~\ref{sec:vortex}. On the real domain $q/\kappa>v_c$,
the flux sum converges and gives the handle factor of the
larger root selected by the vortex sum. In the flux calculation
we chose the expansion in $x/y$. To derive this contour from
the full parent theory, one would also have to include the
coupling $U\log\Phi$ and integrate over the hatted gauge
multiplets.

\section{Critical loci in the quintic hatted theory}
\label{app:geometry}

\subsection{Normal directions and generic critical points}

In the geometric phase, each $b_i$ is linear in $\widehat p$.
Near $\widehat p=G=0$, we can therefore write
\begin{equation}
 w_L=\widehat p\,G(\widehat u)
     +\widehat p^{\,2}H(\widehat p,\widehat u).
\end{equation}
For a smooth quintic, $dG\ne0$ along $G=0$, so we can use
$n'=G+\widehat pH$ as a local coordinate in the normal
direction. In this coordinate, $w_L=\widehat p\,n'$.

In the LG phase, we set $\widehat p=1$ and write the
superpotential in terms of $X_1,\ldots,X_5$, as in
\eqref{eq:W5}. We have already studied the critical point
at $X_1=\cdots=X_5=0$. We now ask whether $w_L$ has any
other interacting critical points.

Suppose that $X_k=0$ for some $k$. For any $j$ with
$X_j\ne0$, the derivative of the product monomial
$b_0=\prod_i X_i$ with respect to $X_j$ vanishes, since
it still contains the factor $X_k$. Thus, away from the
poles of $w_L$, we have
\[
 \partial_{X_j}w_L
 =\frac{5d_jX_j^4}{(1-X_j^5/L)^2}\ne0.
\]
For nonzero $d_1,\ldots,d_5$, the origin is therefore the
only critical point on the coordinate hyperplanes.

Consider now the dense torus, where all the $X_j$ are
nonzero. We regard the five critical equations
$\partial_{X_j}w_L=0$ as functions of both the fields and
the coefficients. Varying $d_1,\ldots,d_5$ at fixed $d_0$
and $L$, we find
\begin{equation}
 \partial_{d_k}\partial_{X_j}w_L
 =\delta_{jk}\frac{5X_j^4}{(1-X_j^5/L)^2},
 \qquad j,k=1,\ldots,5.
\end{equation}
This matrix is invertible on the dense torus away from the
poles, so the joint zero locus in field and coefficient
space is smooth. At a regular value of its projection to the
coefficient space, the five critical equations have full rank
in the five field directions. Their field derivative is the
Hessian of $w_L$. Thus, for generic coefficients, every finite
critical point away from the origin and the poles is Morse.

\subsection{Additional Morse critical points of the reduced superpotential}
\label{app:extra}

Let us give an example in which we can find some of the
critical points away from the origin explicitly. We take
$d_0=2$ and $d_1=\cdots=d_5=-1$, with the remaining
parameters chosen as in \eqref{eq:ray}. For these
coefficients, the homogeneous quintic defines a smooth
hypersurface in $\mathbb P^4$, since
\[
 d_0^5\ne-5^5\prod_{j=1}^5 d_j.
\]

In the LG phase, we set $\widehat p=1$, so that
$b_j=X_j^5$ and $b_0=\prod_j X_j$. Fix a fifth root of
unity $\omega\ne1$ and parameterize the family by $u_\omega=b_1/L$, so that
\[
 b_j=L u_\omega,\qquad j=1,\ldots,5,\qquad b_0=L\omega u_\omega.
\]
For example, we can take $X_1=\omega X_2$ and
$X_2=\cdots=X_5$, with $X_2^5=L u_\omega$.
Along this family, the reduced superpotential is
\begin{equation}
 w_L=L\left(\frac{2\omega u_\omega}{1-\omega u_\omega}
            -\frac{5u_\omega}{1-u_\omega}\right).
\end{equation}
To find critical points of $w_L$ in all five fields, we
evaluate its five logarithmic derivatives on this family.
They all have the same value,
\[
 X_j\partial_{X_j}w_L
 =\frac{2L\omega u_\omega}{(1-\omega u_\omega)^2}
  -\frac{5L u_\omega}{(1-u_\omega)^2}
 \equiv D_\omega(u_\omega).
\]
For $u_\omega\ne0$ and away from the poles, the five critical
equations therefore reduce to
\begin{equation}
 2\omega(1-u_\omega)^2-5(1-\omega u_\omega)^2=0.
 \label{eq:otherroots}
\end{equation}
The discriminant is $40\omega(1-\omega)^2$. Thus, for
each fixed $\omega\ne1$, we obtain two distinct roots.
Neither root is equal to $0$, $1$, $\omega^{-1}$, or $-1$.

We can check that both critical points are Morse by
computing the Hessian in the coordinates $\log X_j$.
In the four directions satisfying
$\sum_j\delta\log X_j=0$, the product $b_0$ is fixed.
The Hessian has the same eigenvalue in all four directions,
\[
 -\frac{25L u_\omega(1+u_\omega)}{(1-u_\omega)^3},
\]
which is nonzero at both roots. The remaining direction
is the common rescaling $X_j\mapsto e^\alpha X_j$.
Under this rescaling, $u_\omega\mapsto e^{5\alpha}u_\omega$, so the
corresponding Hessian eigenvalue is $5u_\omega D_\omega'(u_\omega)$. This is
also nonzero, since both roots are simple.

For each fixed $\omega\ne1$, we have therefore found two
Morse critical points of $w_L$ away from the quintic
critical point at the origin. At both points, the charged
matter in each auxiliary block has nonzero Coulomb mass
on the branch used to derive $w_L$. Whether these critical
points give massive vacua of the full parent theory
remains open.

\section{Vortex normalization and the remaining gauge integral}
\label{app:vortex}

There is a subtlety in taking $x\to0$ in the vortex
calculation. In deriving \eqref{eq:rate}, we took
$\epsilon\to0$ at fixed nonzero $x$. Instead, suppose
that we first set $x=0$ and take $m_H=R_\Phi/2$.
Then $\gamma=1/2$, and the vortex sum simplifies to
\[
 \mathcal V_S(0)
 =\sum_{n=0}^\infty\frac{(\kappa/\epsilon)^n}{n!}
 =e^{\kappa/\epsilon}.
\]
If we first take $\epsilon\to0$ at fixed positive $x$
in the region where the larger root dominates, Stirling's
formula and the Gaussian integral around $y_+$ give
\begin{equation}
 e^{-\mathcal A_x(y_+)/\epsilon}\mathcal V_S
 \longrightarrow
 2\sqrt{\frac{x+y_+}{y_+-x}}
 \qquad (m_H=R_\Phi/2).
\end{equation}
Taking $x\to0$ afterwards, we obtain two rather than one.
Thus, although the exponential rates agree, the prefactor
depends on the order of the two limits.

Another interesting question is what happens if we also
integrate over the $H$ gauge multiplet. For comparison,
consider a reference calculation in which $\Phi$ has a
Gaussian matter action. For $R_\Phi=0$, we choose the
positive $\Phi$ pole together with the $S$ pole. With zero
flavor backgrounds and $q,\kappa>0$, the south-pole vortex
sum is
\begin{equation}
 \sum_{i,j\ge0}
 \frac{(q/\epsilon)^i(\kappa/\epsilon)^j}{i!\,j!}
 \left[\frac{(1/2)_i(1/2)_j}{(1/2)_{i+j}}\right]^2.
 \label{eq:refsum}
\end{equation}
Writing $X=\epsilon i$ and $Y=\epsilon j$, the function
determining the exponential behavior of the sum is
\begin{equation}
 X\log\frac{qX}{(X+Y)^2}
 +Y\log\frac{\kappa Y}{(X+Y)^2}+X+Y.
\end{equation}
For fixed $X+Y$, this is a convex function of
$X/(X+Y)$. Hence, the maximum is attained on one of the
two axes. On the two axes, the maximal values are $q$
and $\kappa$, both of which are larger than the value
at the interior stationary point,
$q\kappa/(q+\kappa)$.

In this example, the dominant saddle in the calculation
at fixed $x$ need not be the dominant saddle of the joint
calculation. Of course, in the parent theory we also
have the logarithmic coupling $U\log\Phi$ and the mixed
gauge interaction in \eqref{eq:MP}. The above reference
calculation does not take these interactions into account.

\section{Quintic gauge groups and the two GLSM moduli in the six-monomial family}
\label{app:integral}

For the six-monomial quintic, the kernel of $P$ gives
the charge vector of the ordinary $U(1)$ gauge group.
To determine the finite gauge group on the mirror side,
we need the integral factorization of $P$.

Following Example~3 of~\cite{MP}, let $\mathsf S$ be
the $6\times5$ matrix consisting of the first five
columns of $P$, labelled by $0,\ldots,4$. The relation
$P\ell=0$ then gives the factorization
\begin{equation}
 \mathbf v=(5,-1,-1,-1,-1)^T,\qquad
 \mathsf T=(I_5\mid\mathbf v),\qquad
 P=\mathsf S\mathsf T.
 \label{eq:appFactorization}
\end{equation}
Since $\mathsf T$ contains the identity matrix $I_5$,
its map on the integral lattices is surjective and has
kernel $\Z\ell$. Thus, on the ordinary side we have
\begin{equation}
 G=\{(\lambda^{-5},\lambda,\ldots,\lambda)
       \mid\lambda\in U(1)\}.
\end{equation}
For a gauge bundle of degree $n$, the flux associated
with the $i$-th field is therefore $n\ell_i$.

On the mirror side, we consider $\ker\mathsf S^T$.
Since $\mathsf T^T$ is injective on the compact tori,
we have $\ker\mathsf S^T=\ker P^T$. If $g_0$ denotes
the phase acting on $\widehat{\mathcal P}$ and $g_j$
the phase acting on $\widehat{\mathcal U}_j$, the
kernel conditions are
\begin{equation}
 g_0\prod_jg_j=1,\qquad g_0g_j^5=1
 \quad(1\leq j\leq5).
 \label{eq:appHatKernel}
\end{equation}
For a given $g_0$, choose $\lambda$ such that
$\lambda^{-5}=g_0$. Writing $\omega_j=g_j/\lambda$,
we have $\omega_j^5=1$ and $\prod_j\omega_j=1$.
Thus, we obtain the surjective map
\begin{equation}
 U(1)\times\Gamma\longrightarrow\widehat G,\qquad
 (\lambda,\omega)\longmapsto
 (\lambda^{-5},\lambda\omega_1,\ldots,\lambda\omega_5),
 \qquad
 \Gamma=\{\omega\in\mu_5^5:\ \textstyle\prod_j\omega_j=1\}.
\end{equation}
The kernel of this map consists of
$(\zeta,(\zeta^{-1},\ldots,\zeta^{-1}))$ with $\zeta\in\mu_5$.

In the geometric phase, the diagonal $\mu_5$ is included
in the projective identification, so that the remaining
finite group is
\[
 \Gamma/\mu_5^{\rm diag}\simeq(\Z/5)^3.
\]
In the LG phase, on the other hand, fixing
$\widehat p=1$ leaves the full group
$\Gamma\simeq(\Z/5)^4$. The diagonal $\mu_5$ should also
be included when we consider the twisted sectors of
the LG orbifold.

For example, on a torus, the sum over $\Gamma$ bundles
is a sum over two holonomies. More generally, on a
connected genus-$g$ Riemann surface, the orbifold
calculation involves a sum over $2g$ holonomies with
the factor $1/|\Gamma|$, assuming trivial discrete
torsion. Indeed, a flat bundle is specified by a homomorphism
from the surface fundamental group to the abelian group
$\Gamma$, and each such bundle has automorphism group
$\Gamma$. Since all $B_i$ are invariant under $\Gamma$,
the column calculation cannot distinguish these
holonomies. We need to recover the bundle sum from
the parent theory.

On the A side, there is no additional finite gauge
group from $K_i$ in the $S$-Higgs vacuum, since each
$S_i$ has charge one under $K_i$.

Finally, the six-monomial subfamily has two algebraic
moduli. Writing
\[
 z(v)=v_0^{-5}\prod_{j=1}^5v_j,
\]
the MP correspondence relates the ordinary FI
coordinate to $z(d)$ and the hatted FI coordinate
to $z(c)$~\cite{MP}. From the threshold calculation,
we have
\[
 z(d)=\frac{z(\kappa)^2}{z(a)}.
\]
With $\kappa_i=Ld_i$ and $a_i=-L^2d_i$, this gives
$z(a)=z(d)$ independently of $L$.

The other parameter $z(c)$ is independent of this choice.
Near the LG point of the hatted GLSM, its FI deformation
corresponds to a twisted-sector marginal operator~\cite{Witten,IV}.
Thus, in comparing the two theories, matching $z(a)$
with $z(d)$ is not sufficient. We also need to specify
the second modulus and the orbifold projection.

\end{document}